\documentclass[12pt]{iopart}
\usepackage{citesort}
\usepackage[T1]{fontenc} 
\usepackage{iopams}  
\begin{document}

\title{Reviving The Shear-Free Perfect Fluid Conjecture In General Relativity}

\author{Muzikayise E. Sikhonde}

\address{Cosmology and Gravity Group,
Department of Mathematics and Applied Mathematics,
University of Cape Town, Rondebosch 7701, Cape Town, South Africa}
\ead{skhmuz002@myuct.ac.za, skhmuz002@gmail.com, sikhome@unisa.ac.za}
\vspace{10pt}
\author{Peter K. S. Dunsby}

\address{Cosmology and Gravity Group,
Department of Mathematics and Applied Mathematics,
University of Cape Town, Rondebosch 7701, Cape Town, South Africa}
\ead{peter.dunsby@uct.ac.za}
\vspace{10pt}

\begin{indented}
\item[]\today{}
\end{indented}

\begin{abstract}
Employing a Mathematica symbolic computer algebra package called xTensor, we present $(1+3)$-covariant special case proofs of the shear-free perfect fluid  conjecture in General Relativity. We first present the case where the pressure is constant, and where the acceleration is parallel to the vorticity vector. These cases were first presented in their covariant form by Senovilla et. al.  We then provide a covariant proof for the case where the acceleration and vorticity vectors are orthogonal, which leads to the existence of a Killing vector along the vorticity. This Killing vector satisfies the new constraint equations resulting from the vanishing of the shear. Furthermore, it is shown that in order for the conjecture to be true, this Killing vector must have a vanishing spatially projected directional covariant derivative along the velocity vector field. This in turn implies the existence of another \textit{basic} vector field along the direction of the vorticity for the conjecture to hold. Finally, we show that in general, there exist a \textit{basic} vector field parallel to the acceleration for which the conjecture is true.
\end{abstract}

%
%
%
%
%
\section{Introduction}
\label{Sec1}

This paper presents a method of investigating shear-free perfect fluid solutions in General Relativity. This method is used to examine a number of special cases. The shear-free perfect fluid conjecture in General Relativity states that, {\it a shear-free velocity vector field of a barotropic perfect fluid with $\mu+p\neq 0$ and  $p=p({\mu})$, is either expansion or rotation free} \cite{senovilla1998theorems}, where $\mu$ and $p$ are the energy density and pressure of the fluid respectively. The first indication that the vanishing of the shear can have restrictive properties between the rotation and the expansion of a cosmological model was given by G$\ddot{\text{o}}$del \cite{godel1952rotating,godel1949example}. Some of the well known examples of these cosmological models include: the Einstein (static) universe, FLRW universes (expanding models) and the G$\ddot{\text{o}}$del (purely rotating) universe. \newline

The requirement that the matter equation of state be of a barotropic perfect fluid produces new additional constraints on the full Einstein field equations, for instance all expanding and non-rotating shear-free perfect fluids with a barotropic equation of state are known to exist \cite{collins1986shear}. In contrast, not all non-expanding and rotating shear-free perfect fluids with a barotropic equation of state are known \cite{karimian2012contributions}, but a fair amount of these models are known to exist especially all stiff rotating axisymmetric and stationary perfect fluids belong in this class. A simultaneously expanding and rotating model (Bianchi IX) was found to have an equation of state of the form $p=-\mu=\text{constant}$, which is of the cosmological constant type and is excluded in this discussion \cite{obukhov2002shear}.\newline

The validity of the above mentioned conjecture would be a striking feature of the full Einstein field equations, while in Newtonian cosmology there exist shear-free perfect fluids with a barotropic equation of state which are both expanding and rotating simultaneously \cite{ellis2011shear,heckmann1955bemerkungen,narlikar1963newtonian,senovilla1998theorems}. Whereas in the $f(R)$ theory of gravity, there exist no counterpart of the conjecture \cite{sofuouglu2014investigations}, and also where the equations of $f(R)$ gravity are linearized about the FLRW backgound, the counterpart does not exist as well \cite{abebe2011shear}. It also turns out that the conjecture is a feature of the Lorenzian signature because in the Riemannian signature one can obtain a {\it Ricci-flat} metric where shear-free, expanding and rotating solution exist \cite{pantilie1999harmonic}. \newline

Following G$\ddot{\text{o}}$del's assertion in \cite{godel1952rotating}, Sch$\ddot{\text{u}}$cking in \cite{schucking1957homogene}  gave a more precise statement which was followed by a coordinate based proof of the conjecture for spatially homogeneous dust($p=0$) cosmological models. Banerji \cite{banerji1968homogeneous} considered the situation for non-zero pressure, and they also provided a coordinate based proof of the conjecture for {\it tilted} spatially homogeneous perfect fluid solution with $\gamma$-law equation of state $p=(\gamma-1)\mu$ ,where $\gamma-1 \neq \tfrac{1}{9}$. In 1967 Ellis \cite{ellis1967dynamics} provided a generalization of the result obtained by Sch$\ddot{\text{u}}$cking, by employing the orthogonal tetrad formalism to prove that one does not need to impose the spatial homogeneity condition in dust models for the conjecture to be true. This result was shown to hold even in the presence of the cosmological constant by White and Collins in 1984 \cite{white1984class}.\newline

The case for incoherent radiation, where the equation of state is $p=\tfrac{1}{3}\mu$ was proven by Treciokas and Ellis \cite{treciokas1971isotropic}, also using an orthonormal tetrad formalism together with a specific coordinate system. Coley showed that the conjecture remains true for incoherent radiation with a cosmological constant \cite{coley1991fluid}. A sketch of a proof for perfect fluids where the derivative along the fluid four-velocity of the potential for the acceleration $r=\int^{p}_{p_{0}}\frac{1}{p+\mu}dp$, obeys the following equation $\dot{r}=\beta(r)$ was provided in \cite{treciokas1971isotropic}. For spatially homogeneous spacetimes this result implies that the conjecture remains true for a general equation state \cite{banerji1968homogeneous,king1973tilted,white1984class}. \newline

Lang and Collins \cite{lang1988observationally,Lang:1993:CSG:920627} provided a detailed proof of the above result, where they showed that when $\Theta=\Theta(\mu)\footnote{Where by the conservation equation $\dot{\mu}=-(\mu+p)\Theta$, implies $\dot{r}=\beta(r)$.}$, then it follows that $\Theta\omega=0$, where $\Theta$ and $\omega$ are the expansion and vorticity scalars respectively. Sopuerta in \cite{Sopuerta:1998yc} gave a fully covariant proof of this results. After a succession of papers that proved the conjecture where the acceleration vector is parallel to the vorticity \cite{white1984class,senovilla1998theorems} or where the magnetic or the electric parts of the Weyl tensor vanishes \cite{collins1984shear,cyganowski2000shear,Lang:1993:CSG:920627}, Collins \cite{collins1986shear} conjectured that shear-free perfect fluids which rotate and expand do not exist in general, except in very special conditions. This conjecture is still yet to be proven.\newline

In search for a full proof of this conjecture, some authors have shown that the conjucture is true in a wide range of special conditions; namely for the cases where the speed of sound $\tfrac{dp}{d\mu}=-\tfrac{1}{3}$ \cite{cyganowski2000shear,Lang:1993:CSG:920627,slobodeanu2014harmonic}, where $\Theta=\Theta(\omega)$ \cite{Sopuerta:1998yc}; Petrov type N and III in \cite{carminati1990shear} and \cite{carminati1996shear,carminati1997shear} respectively, a case where there exist a conformal killing vector parallel to the velocity vector \cite{coley1991fluid}. Also in the case of a divergence-free electric part of the Weyl tensor \cite{van2012shear}, or the case with a $\gamma$-law equation of state where the magnetic part of the Weyl tensor is divergence-free \cite{van2007shear}, and one where the equation of state is more generic \cite{Carminati:2009zz}. Nzioki, Goswami, Dunsby and Ellis proved the case in which the Einstein field equations are linearized with respect to the FLRW background \cite{nzioki2011shear}.\newline

Recently, Slobodeanu \cite{slobodeanu2014shear} provided a proof of the conjecture for a $\gamma$-law equation of state with a zero cosmological constant which excludes the following cases ($\gamma-1=-\tfrac{1}{5},-\tfrac{1}{6},-\tfrac{1}{11},-\tfrac{1}{21},\tfrac{1}{15},\tfrac{1}{4}$). Carminati \cite{carminati2015shear} attempted to prove the conjecture for a linear equation of state with a zero cosmological constant, but the proof was incorrect since there was an in appropriate use of a differential equation solving command in Maple. Finally, Van den Bergh and Slobodeanu in \cite{van2016shear} completed the proof attempted by Carminati in \cite{carminati2015shear}, thus proving the conjecture for the case of a linear equation of state , including a non-vanishing cosmological constant. In this work they reduced the problem to a Lemma, which states that if a rotating and expanding shear-free perfect fluid, obeying a linear equation of state  $p=(\gamma-1)\mu+constant$ together with some defined {\it basic} variables, then a Killing vector exists along the vorticity. And they claim that this lemma is valid even for a general barotropic equation of state. \newline

In the present paper, we first presented a much more compact and clearer covariant proof of the shear-free conjecture in the case where the pressure is constant,  followed by the proof of the case where the acceleration and the vorticity vector are parallel. Both these cases were dealt with in the presence of a non-vanishing cosmological constant. These two cases were first presented by Senovilla et. al. \cite{senovilla1998theorems} in their covariant form. We then provide a covariant proof which was first given in \cite{van2016shear} using the orthonormal tetrad approach for the case where the acceleration and the vorticity vectors are orthogonal. \newline

The above mentioned proof shows that whenever the acceleration is orthogonal to the vorticity vector, there is a Killing vector along the vorticity under which the constraints resulting from the vanishing of the shear are satisfied and remain so under time-propagation. With further analysis, we show that this Killing vector must have the property that its spatially projected covariant directional derivative along $u^{a}$ must vanish in order for the conjecture to be true, which essentially means that there exist another vector field along the direction of the vorticity with a vanishing spatially projected Lie derivative along the velocity, implying that it is \textit{basic}. Finally, we show that in general if the acceleration is non-zero, there must exist a vector field along the direction of the acceleration, with a vanishing spatially projected Lie derivative along the velocity in order for the conjecture to be true. \newline

This paper is organized as follows, In section (\ref{Sec2}) we discuss the various method used to check for consistency in General Relativity. In section (\ref{Sec3}) we introduce the $(1+3)$ covariant dynamical equations and constraints equations of general relativity, and then we discuss the propagation of the constraints along fluid flow lines. In section (\ref{Sec4}) we present the shear-free $(1+3)$ covariant dynamical equations and constraints equations of general relativity. In section (\ref{Sec5}) we propagate the new shear-free constraints in order to obtain new consistency relations. In section (\ref{Sec6}) we present the proof of the shear-free perfect fluid theorem in general relativity in the dust case $(p=0)$, followed by the discussion of the conjecture for the general equation of state, where we have produced new 13 levels of constraint equations, which must be checked for consistency.\newline

Then in section (\ref{Sec7}) we revisit the proof of the theorem where the vorticity is parallel to the acceleration vector. In section (\ref{Sec8}) we discuss the proof that the orthogonality of the acceleration and the vorticity vectors leads us to the existence of a Killing vector along the vorticity, and we present a proof that this Killing vector has to have it's directional covariant derivative along the velocity equal to zero in order for the theorem to be true. And in section (\ref{Sec9}) we show that this is equivalent to the existence of a \textit{basic} vector field along the vorticity, which means its spatially projected Lie derivative along the velocity vanishes. In section (\ref{Sec10}) we show that if the acceleration is non-zero, there exist a \textit{basic} vector filed along it, for the conjecture to be true. In section (\ref{Sec11}) we present our conclusions and we discuss the future prospects for this endeavor. Finally, we included some useful equations and identities in the appendix (\ref{Appendix}). 
\section{The consistency check methods}\label{Sec2}
\subsection{Testing consistency in General Relativity}
In general, the ten Einstein gravitational field equations (EFE)\footnote{$R_{ab}-\tfrac{1}{2}g_{ab}R=T_{ab}-\Lambda g_{ab}.$} are fully consistent. This is so, provided no geometric restrictions are imposed and the energy-momentum tensor (EMT) is conserved\footnote{Meaning $\nabla_{a}T^{ab}=0$, where $T^{ab}$ is the EMT.}. The EFE are divided into four constraint equations together with six evolution equations, as a consequence of the EMT being conserved, the spacetime unfolds uniquely from initial data satisfying the four constraint equation. However, since the solutions are subject to general covariance (GC) allowed by General Relativity, thus they can be expressed in many different forms for the same geometrical scenario. Hence, essentially the issue is, what aspect of the solution represent coordinate or physical degrees of freedom? \\
\newline
Whenever some geometric or physical constrains are imposed on the EFE, for an example the restriction of matter to a perfect fluid or the invariance of spacetime under some group symmetries, then the consistency of the EFE is no longer necessarily true. This has to be investigated on a case by case basis. Essentially, imposing these constraints on the EFE leads to new constraint (NC) equations \footnote{Of the form $F(g_{ab},T_{ab})=0$.} which may arise from the conversion of a propagation equation into a constraint equation.  It follows that one has to take the time derivative of the NC and using the propagation equations together with the original constraint equations and the NC itself for substitution, one obtains the next level of new constraint equations. These will either be identically satisfied, and consistency has been shown, or inconsistent, in which case the solution does not exist, or neither, in which case one has to time propagate the previously obtained constraint equations and repeat the substitution procedure to get further constraints and so on, until one has either shown consistency or inconsistency.  \\
\newline
It may sometimes be useful to use the Extended Einstein Field Equations (EEFE), where one adjoins to the metric and matter variables the components of the Weyl tensor\footnote{$C_{abcd}$ is the trace-free part of $R_{abcd}$} and the extra consistency conditions\footnote{$R_{ab[cd;e]} = 0$} which in effect act as field equations for the Weyl tensor by determining its divergence, with the EMT as the source term. Under the extra consistency conditions, and the spacetime being algebraically special, one can investigate the resulting consistency conditions as discussed above. The difference is simply that one must add the extra evolution and constraint equations from the electric and the magnetic parts of the Weyl curvature tensor. 
\subsection{The different formalisms for testing consistency}
The consistency conditions can be investigated as discussed above in terms of the following formalisms,
\begin{itemize}
\item[(a)] A coordinate representation, which is the traditional way. However one then has to carefully keep track of the coordinate freedom consistent with the geometry considered, and try thereby to distinguish physical from coordinate effects.
\item[(b)] A tetrad representation, where one adds as auxiliary variables the tetrad rotation coefficients, which will have physical meaning if the tetrad vectors are chosen in a physically significant way. This can be done either for a null tetrad, as per Newman and Penrose (where the Weyl tensor components are added as extra variables) or the closely related spinor formalism, or for an orthonormal tetrad, as per Ellis \cite{ellis1967dynamics} and Ellis and MacCallum \cite{ellis1969class}. The former is better for studies of gravitational waves and vacuum solutions, and the latter for studies of fluid solutions, and possibly kinetic theory solutions.
\item[(c)] A (1+3) covariant formalism, as per Ehlers et. al~\cite{ehlers1961beitrage}, Kristian and Sachs~\cite{kristian1966observations}, Tr$\ddot{\text{u}}$mper \cite{trumper1965special}, and Hawking \cite{hawking1967occurrence}, as summarized by Ellis \cite{ellis1971topology} and Van Elst and Ellis \cite{ellis1999cosmological}. In this case it is convenient to express the Weyl tensor in terms of its electric\footnote{$E_{ab}$ symmetric trace-free tensor} and magnetic\footnote{$H_{ab}$ symmetric trace-free tensor} parts relative to the chosen 4-velocity vector. Then there are propagation and constraint equations for the electric and magnetic Weyl tensor parts deriving from the consistency relations that are imposed by the geometric or physic al restrictions.   The full set of equations are laid out most comprehensively in Ellis and Van Elst \cite{ellis1999cosmological}. They can be investigated covariantly as far as possible, and then developed in terms either of a coordinate basis or a tetrad frame, as seems appropriate. This is the formalism employed in this work.
\end{itemize}
\section{The (1+3)-covariant dynamical equations}\label{Sec3}
\subsection{Preliminaries}
Units: $c=1=8\pi G/c^2$.
\paragraph{Assumption:}  On a cosmological space-time manifold $(\mathcal{M},\bold{g})$, where $\bold{g}$ or in index form $g_{ab}$ is the metric tensor for the manifold $\mathcal{M}$, there exist a future directed, normalised, timelike velocity vector field $\bold{u}$ or in index form $u_{a}$, which describes a preferred reference geodesic (a congruence) in an entire system of nonintersecting geodesics. The scenario $(\mathcal{M},\bold{g},\bold{u})$ forms the basis for a relativistic cosmological model, which is subject to the Einstein field equations below~\cite{ellis1999cosmological}, 
\begin{align}
\label{EFE}
G_{ab}&=T_{ab}-\Lambda g_{ab}.
\end{align}
In equation \eqref{EFE} above, $G_{ab}=R_{ab}-\tfrac{1}{2}g_{ab}R$ is the Einstein tensor. $G_{ab}$ is constructed from the first and the second contractions of the Riemann curvature tensor with the metric tensor. Where $R_{ab}=R^{c}{}_{acb}$ and $R=R^{c}{}_{ac}{}^{a}$ are the Ricci tensor and scalar respectively.
$T_{ab}$ is the energy-momentum tensor and $\Lambda$ is the cosmological constant (where $T_{ab}$ and $\Lambda$ have the property that $\nabla_{a}T^{ab}=0$ and $\nabla_{a}\Lambda=0$).

\paragraph{(1+3)-projection tensors:} 
\begin{subequations}\label{eq:PT}
\begin{align}
\label{eq:PT:1}
U^{a}{}_{b}=-u^{a}u_{b} \hspace{.1 in}, \hspace{.1 in} U^{a}{}_{a}=1, \hspace{.1 in} h^{a}{}_{b}=\delta^{a}{}_{b}-U^{a}{}_{b} \hspace{.1 in} \text{and} \hspace{.1 in} h^{a}{}_{a}=3
\end{align}
\end{subequations}

\paragraph{(1+3)-projected covariant derivatives and tensors:} 
\begin{subequations}\label{eq:PCD}
The covariant time derivative along $\bf u$ is given by the $\bf \dot{}$ operator in equation \eqref{eq:PCD:1} below, and the fully orthogonally projected covariant derivative is given by the $\tilde{\nabla}$ operator in equation \eqref{eq:PCD:2} below for any tensor $T^{ab}{}_{cd}$ respectively. We will use the angle brackets in the indices of tensors to indicate the orthogonally projected symmetric trace-free part, whereas for vectors the angle brackets of the indices indicates orthogonal projection (as follows, $\dot{T}^{\langle ab\rangle}=\left[h^{(a}{}_{f}h^{b)}{}_{g}-\tfrac{1}{3}h^{ab}h_{fg}\right]\dot{T}^{fg}$ and $\dot{v}^{\langle a\rangle}=h^{a}{}_{f}\dot{v}^{f}$).
\begin{align}
\label{eq:PCD:1}
\dot{T}^{ab}{}_{cd}&=u^{e}\nabla_{e}T^{ab}{}_{cd}
\\
\label{eq:PCD:2}
\tilde{\nabla}_{e}T^{ab}{}_{cd}&=h^{a}{}_{f}h^{b}{}_{g}h^{p}{}_{c}h^{q}{}_{d}h^{r}{}_{e}\nabla_{r}T^{fg}{}_{pq}
\end{align}
\end{subequations}

\paragraph{3 - volume element:} 
\begin{subequations}\label{eq:VE}
In equation \eqref{eq:VE:1} $\epsilon_{abcd}$ is the spacetime 4-volume element with the following property $(\epsilon_{abcd}=\epsilon_{[abcd]}, \epsilon_{0123}=\sqrt{|\text{det} g_{ab}|})$. Giving us the three volume element below, 
\begin{align}
\label{eq:VE:1}
\epsilon_{abc}=u^{g}\epsilon_{gdef}h^{d}{}_{a}h^{e}{}_{b}h^{f}{}_{c} .
\end{align}
\end{subequations}

\paragraph{The covariant derivative of $\bf u$:} 
\begin{subequations}\label{eq:CDuG}
In equation \eqref{eq:CDuG:1} the covariant derivative of $u_{a}$ has been decomposed into its irreducible kinematic quantities described in equation \eqref{eq:CDuG:2}, which are the acceleration vector $\dot{u}_{a}$, the expansion scalar $\Theta$, the shear tensor $\sigma_{ab}$, and the vorticity vector $\omega_{a}$, 
\begin{align}
\label{eq:CDuG:1}
\nabla_{a}u_{b}&=-u_{a}\dot{u}_{b}+\tfrac{1}{3}\Theta h_{ab}+\sigma_{ab}+\epsilon_{abc}\omega^{c}
\\
\label{eq:CDuG:2}
\dot{u}^{a}:=u^{b}\nabla_{b}u^{a} \hspace{.1 in} , \hspace{.1 in} \Theta&:=\nabla_{a}u^{a} \hspace{.1 in}, \hspace{.1 in} \sigma_{ab}:=\tilde{\nabla}_{\langle a}u_{b\rangle} \hspace{.1 in}, \hspace{.1 in} \omega_{a}:=\tfrac{1}{2}\epsilon_{abc}\tilde{\nabla}^{b}u^{c}.
\end{align}
\end{subequations}

\paragraph{The energy-momentum tensor $T_{ab}$:} 
\begin{subequations}\label{eq:EMTG}
In equation \eqref{eq:EMTG:1} the energy-momentum tensor $T_{ab}$ has been decomposed into its irreducible matter variables described in equation \eqref{eq:EMTG:2}, which are the energy density $\mu$, the pressure $p$, the momentum density $q^{a}$, and the anisotropic pressure $\pi_{ab}$. 
\begin{align}
\label{eq:EMTG:1}
T_{ab}&=\mu u_{a}u_{b} + 2q_{(a}u_{b)} + p h_{ab} + \pi_{ab}
\\
\label{eq:EMTG:2}
\mu:=T_{ab}u^{a}u^{b} \hspace{.1 in} , \hspace{.1 in} p&:=\tfrac{1}{3}T_{ab}h^{ab} \hspace{.1 in}, \hspace{.1 in} q^{a}:=-T_{cb}h^{ca}u^{b} \hspace{.1 in}, \hspace{.1 in} \pi_{ab}:=T_{cd}h^{c}{}_{\langle a}h^{d}{}_{b\rangle}
\end{align}
\end{subequations}

\paragraph{The Riemann curvature tensor:} 
\begin{subequations}\label{eq:RCTG}
The Riemann curvature tensor $R^{ab}{}_{cd}$ from which the Einstein tensor $G_{ab}$ in equation \eqref{EFE} is constructed, can be decomposed as seen in equation \eqref{eq:RCTG:1} below, where in equation \eqref{eq:RCTG:2} $E_{ab}$ and $H_{ab}$ are the electric and the magnetic parts of the Weyl curvature tensor $C^{ab}{}_{cd}$ respectively, and the Weyl tensor is the trace-less component of the Riemann tensor.
\begin{align}
\label{eq:RCTG:1}
R^{ab}{}_{cd}&=4u^{[a}u_{[c}(E^{b]}{}_{d]}-\tfrac{1}{2}\pi^{b]}{}_{d]})+4h^{[a}{}_{[c}(E^{b]}{}_{d]}+\tfrac{1}{2}\pi^{b]}{}_{d]})\nonumber\\
&\qquad +\tfrac{2}{3}(\mu+3p-2\Lambda)u^{[a}u_{[c}h^{b]}{}_{d} +\tfrac{2}{3}(\mu+\Lambda)h^{[a}{}_{[c}h^{b]}{}_{d}\nonumber\\
&\qquad +2\epsilon^{abe}u_{[c}(H_{d]e}+\tfrac{1}{2}\epsilon_{d]ef}q^{f})+2\epsilon_{cde}u^{[a}(H^{b]e}+\tfrac{1}{2}\epsilon^{b]ef}q_{f})
\\\nonumber\\
\label{eq:RCTG:2}
E_{ab}&:=C_{cdef}h^{c}{}_{a}u^{d}h^{e}{}_{b}u^{f} \hspace{.1 in} , \hspace{.1 in} H_{ab}:=-\tfrac{1}{2}\epsilon_{cdgh}C^{gh}{}_{ef}h^{c}{}_{a}u^{d}h^{e}{}_{b}u^{f}.
\end{align}
\end{subequations}

\paragraph{The vorticity scalar:} 
\begin{subequations}\label{eq:VSG}
The vorticity scalar is constructed from the vorticity vector in the following way,
\begin{align}
\label{eq:VSG:1}
\omega^{2}&=\omega_{a}\omega^{a}\geq 0.
\end{align}
\end{subequations}
\paragraph{The shear scalar:} 
\begin{subequations}\label{eq:SSG}
The vorticity scalar is constructed from the vorticity vector in the following way,
\begin{align}
\label{eq:SSG:1}
\sigma^{2}&=\tfrac{1}{2}\sigma_{ab}\sigma^{ab}\geq 0.
\end{align}
\end{subequations}
\subsection{Ricci identities}
The following set of equations arises from the Ricci identity for $\bf u$, i.e.
\begin{equation}
2\nabla_{[a}\nabla_{b]}u^{c}=R_{ab}{}^{c}{}_{d}u^{d}\label{RicciIDG}
\end{equation}
by using \eqref{eq:CDuG:1} and \eqref{eq:RCTG:1}.
\subsubsection{Time derivative equations:}
Taking the trace of the parallel(to ${\bf u}$) projection of equation \eqref{RicciIDG}, we obtain the Raychaudhuri equation \eqref{eq:RIG:1} which describes gravitational attraction and demonstrates the repulsive property of a positive cosmological constant($\Lambda$). Then by taking the orthogonally projected symmetric trace free part of equation \eqref{RicciIDG}, we obtain the shear propagation equation \eqref{eq:RIG:2} which is crucial in the discussion of shear-free perfect fluids. The orthogonally projected antisymmetric part of the parallel(to ${\bf u}$) projection of equation \eqref{RicciIDG} gives us the vorticity propagation equation \eqref{eq:RIG:3} below,
\begin{subequations}\label{eq:RIG}
\begin{align}
\label{eq:RIG:1}
\dot{\Theta} -\tilde{\nabla}_{a}\dot{u}^{a}&= - \tfrac{1}{3} \Theta^2 + \dot{u}_a\dot{u}^a - 2 (\sigma^2 - \omega^2) -  \tfrac{1}{2} (\mu +  3 \mathit{p})  + \Lambda
\\
\label{eq:RIG:2}
\dot{\sigma}^{\langle ab\rangle} -\tilde{\nabla}^{\langle a}\dot{u}^{b\rangle}&= - \tfrac{2}{3} \Theta \sigma^{ab} + \dot{u}^{\langle a}\dot{u}^{b\rangle} - \sigma^{\langle a}{}_c\sigma^{b\rangle c} - \omega^{\langle a} \omega^{b\rangle} - (E^{ab} - \tfrac{1}{2} \pi^{ab})
\\
\label{eq:RIG:3}
\dot{\omega}^{\langle a\rangle} -\tfrac{1}{2} \epsilon^{abc} \tilde{\nabla}_{b}\dot{u}_{c}&= - \tfrac{2}{3} \Theta \omega^{a} +\sigma^{a}{}_{b}\omega^{b}.
\end{align}
\end{subequations}

\subsubsection{Constraint equations:}
Orthogonally projecting equation \eqref{RicciIDG} and contracting $b$ and $c$, one obtains constraint equation \eqref{eq:RICEG:1} and by contracting with $\epsilon^{ab}{}_{c}$ we obtain the vorticity divergence equation \eqref{eq:RICEG:2}, when we contract with $\epsilon^{abe}$ then taking the symmetric trace-free part of the resulting equation we arrive at the constraint equation for the magnetic part of the Weyl tensor $H^{ab}$ in equation \eqref{eq:RICEG:3}, 
\begin{subequations}\label{eq:RICEG}
\begin{align}
\label{eq:RICEG:1}
0&=(C_1)^{a} := \tilde{\nabla}_{b}\sigma^{ab} - \tfrac{2}{3} \tilde{\nabla}^{a}\Theta + \epsilon^{a}{}_{bc} \left(\tilde{\nabla}^{b}\omega^{c}+2 \dot{u}^{b} \omega^{c}\right) + q^{a}
\\
\label{eq:RICEG:2}
0&=(C_2) :=  \tilde{\nabla}_{a}\omega^{a}- \dot{u}_{a} \omega^{a} 
\\
\label{eq:RICEG:3}
0&=(C_3)^{ab} := H^{ab} + 2\dot{u}^{\langle a} \omega^{b\rangle} + \tilde{\nabla}^{\langle a}\omega^{b\rangle} -\epsilon^{cd\langle a}\tilde{\nabla}_{c}\sigma^{b\rangle}{}_{d}.
\end{align}
\end{subequations}

\subsection{(Contracted) second Bianchi identities}
The following equations are derived from the Bianchi identities by applying equations \eqref{eq:CDuG:1} and \eqref{eq:RCTG:1} in equation \eqref{BianchiIDG} below,
\begin{equation}
\nabla_{[a}R_{bc]d}{}^{e}=0.\label{BianchiIDG}
\end{equation}
\subsubsection{Time derivative equations:}
Contracting $a$ and $e$ in equation \eqref{BianchiIDG} and projecting parallel(to $\bf u$), we obtain two propagation equations below, one for the electric part \eqref{eq:BIG:1} and the other for the magnetic part \eqref{eq:BIG:2} of the Weyl tensor, where these two equations describes gravitational radiation i.e they can be combined to form wave equations for $E^{ab}$ and $H^{ab}$. Then by contracting twice the Bianchi identities \eqref{BianchiIDG}, we are lead to the following energy momentum conservation equation $\nabla_{a}T^{ab}=0$, for which if we project along $\bf u$ we obtain the energy conservation equation \eqref{eq:BIG:4}, and if we project orthogonal to $\bf u$, we obtain the momentum conservation equation \eqref{eq:BIG:3},
\begin{subequations}\label{eq:BIG}
\begin{align}
\label{eq:BIG:1}
(\dot{E}^{\langle ab\rangle}+\tfrac{1}{2}\dot{\pi}^{\langle ab\rangle}) - \epsilon^{cd\langle a} \tilde{\nabla}_{c}H^{b\rangle}{}_{d} - \tfrac{1}{2}\tilde{\nabla}^{\langle a}q^{b\rangle}  &= - \tfrac{1}{2}(p +  \mu)\sigma^{ab} -  \Theta (E^{ab} + \tfrac{1}{6}\pi^{ab}) \nonumber\\
&\qquad + 3\sigma^{\langle a}{}_{c}(E^{b\rangle c} - \tfrac{1}{6}\pi^{b\rangle c}) - \dot{u}^{\langle a}q^{b\rangle}\nonumber\\
&\qquad + \epsilon^{cd\langle a}\big[ 2\dot{u}_{c} H^{b\rangle}{}_{d}  +  \omega_{c}(E^{b\rangle}{}_{d}\nonumber\\
&\qquad +\tfrac{1}{2}\pi^{b\rangle}{}_{d})\big]
\\
\label{eq:BIG:2}
\dot{H}^{\langle ab\rangle} + \epsilon^{cd\langle a} \tilde{\nabla}_{c}(E^{b\rangle}{}_{d}-\tfrac{1}{2}\pi^{b\rangle}{}_{d}) &= -  \Theta H^{ab} + 3\sigma^{\langle a}{}_{c}H^{b\rangle c} + \tfrac{3}{2}\omega^{\langle a}q^{b\rangle}  \nonumber\\
&\qquad - \epsilon^{cd\langle a}\big[ 2\dot{u}_{c} E^{b\rangle}{}_{d} -\tfrac{2}{2}\sigma^{b\rangle}{}_{c}q_{d} \nonumber\\
&\qquad -  \omega_{c}H^{b\rangle}{}_{d}\big]
\\
\label{eq:BIG:3}
\dot{q}^{\langle a\rangle}+\tilde{\nabla}^{a}\mathit{p} + \tilde{\nabla}_{b}\pi^{ab}  &= -\tfrac{4}{3}\Theta q^{a} - \sigma^{a}{}_{b}q^{b}-\left(\mathit{p} + \mu\right)\dot{u}^{a} \nonumber\\
&\qquad -\dot{u}_{b}\pi^{ab}-\epsilon^{abc}\omega_{b}q_{c}
\\
\label{eq:BIG:4}
\dot{\mu}+\tilde{\nabla}_{a}q^{a} &= -\Theta \left( p +  \mu\right)-2\dot{u}_{a}q^{a}-\sigma_{ab}\pi^{ab}.
\end{align}
\end{subequations}

\subsubsection{Constraint equations:}
Projecting the once contracted equation \eqref{BianchiIDG} orthogonal to $\bf u$, we  obtain the two constraints equations \eqref{eq:BICEG:1} and \eqref{eq:BICEG:2} for the divergence of the electric(which is sourced by the spatial gradient of the energy density) and magnetic(which is sourced by the vorticity vector) parts of the Weyl curvature tensor below,
\begin{subequations}\label{eq:BICEG}
\begin{align}
\label{eq:BICEG:1}
0&=(C_4)^{a} := \tilde{\nabla}_{b}(E^{ab}+\tfrac{1}{2}\pi^{ab}) - \tfrac{1}{3} \tilde{\nabla}^{a}\mu -\tfrac{1}{3}\Theta q^{a} - \tfrac{1}{2}\sigma^{a}{}_{b}q^{b} - 3 H^{a}{}_{b} \omega^{b}\nonumber\\
&\qquad -\epsilon^{abc}\left[\sigma_{bd}H^{d}{}_{c}-\tfrac{3}{2}\omega_{b}q_{c}\right] 
\\
\label{eq:BICEG:2}
0&=(C_5)^{a} := \tilde{\nabla}_{b}H^{ab} + (p  + \mu )\omega^{a} + 3\omega^{b} (E^{a}{}_{b}- \tfrac{1}{6}\pi^{a}{}_{b}) + \epsilon^{abc}\big[\tfrac{1}{2}\tilde{\nabla}_{b}q_{c}  \nonumber\\
&\qquad + \sigma_{bd}(E^{d}{}_{c}+\tfrac{1}{2}\pi^{d}{}_{c})\big].
\end{align}
\end{subequations}
\subsection{Propagation of the constraints along ${\bf u}$}
The time propagation of the constraints $(C_{1})^{a}-(C_{5})^{a}$ along ${\bf u}$, has been achieved through the application of the commutation relations given by van Elst in reference \cite{van1996extensions}, Where the modifications pointed out by MacCallum in reference \cite{Mac98a} were taken into account, these are given as follows,
\begin{align}
(\dot{C}_{1})^{\langle a\rangle} &= - \Theta (C_{1})^{a} - \tfrac{3}{2}\sigma^{a}{}_{b}(C_{1})^{b} + \tfrac{1}{2}\epsilon^{abc}\omega_{b}(C_{1})_{c}-\tfrac{8}{3}\omega^{a}(C_{2}) - \epsilon^{abc}\sigma_{bd}(C_{3})_{c}{}^{d} \nonumber\\
&\qquad - 3\omega_{b}(C_{3})^{ab} - (C_{4})^{a},
\end{align}
\begin{equation}
(\dot{C}_{2}) = - \Theta (C_{2}),
\end{equation}
\begin{align}
(\dot{C}_{3})^{\langle ab\rangle} &= - \Theta (C_{3})^{ab} + 3\sigma^{\langle a}{}_{c}(C_{3})^{b\rangle c} + \epsilon^{cd\langle a}\omega_{c}(C_{3})^{b\rangle}{}_{d} + \tfrac{1}{2}\epsilon^{cd\langle a}\sigma^{b\rangle}{}_{c}(C_{1})_{d} \nonumber\\
&\qquad + \tfrac{3}{2}\omega^{\langle a}(C_{1})^{b\rangle} ,
\end{align}
\begin{align}
(\dot{C}_{4})^{\langle a\rangle} - \tfrac{1}{2}\epsilon^{abc}\tilde{\nabla}_{b}(C_{5})_{c} &= - \tfrac{4}{3}\Theta (C_{4})^{a} + \tfrac{1}{2}\sigma^{a}{}_{b}(C_{4})^{b} - \tfrac{1}{2}\epsilon^{abc}\omega_{b}(C_{4})_{c}-\tfrac{1}{2}(\mu + \mathit{p})(C_{1})^{a} \nonumber\\
&\qquad  - \tfrac{1}{2}\pi^{a}{}_{b}(C_{1})^{b} + 2 \epsilon^{abc}E_{bd}(C_{3})_{c}{}^{d} + \tfrac{3}{2}\epsilon^{abc}\dot{u}_{b}(C_{5})_{c},
\end{align}
\begin{align}
(\dot{C}_{5})^{\langle a\rangle} + \tfrac{1}{2}\epsilon^{abc}\tilde{\nabla}_{b}(C_{4})_{c} &= - \tfrac{4}{3}\Theta (C_{5})^{a} + \tfrac{1}{2}\sigma^{a}{}_{b}(C_{5})^{b} - \tfrac{1}{2}\epsilon^{abc}\omega_{b}(C_{5})_{c} - \tfrac{1}{2}\epsilon^{abc}q_{b}(C_{1})_{c} \nonumber\\
&\qquad  + \tfrac{3}{2}q^{a}(C_{2}) + 2 \epsilon^{abc}H_{bd}(C_{3})_{c}{}^{d} - \tfrac{3}{2}\epsilon^{abc}\dot{u}_{b}(C_{4})_{c}.
\end{align}
\section{The shear-free perfect fluid equations}\label{Sec4}
When we restrict the equations in section (\ref{Sec3}) above to a perfect fluid and the vanishing of the shear, we obtain the following equations,
\paragraph{The covariant derivative of $\bf u$:} 
\begin{subequations}\label{eq:CDu}
\begin{align}
\label{eq:CDu:1}
\nabla_{a}u_{b}&=-u_{a}\dot{u}_{b}+\tfrac{1}{3}\Theta h_{ab}+\epsilon_{abc}\omega^{c}
\\
\label{eq:CDu:2}
\dot{u}^{a}:=u^{b}\nabla_{b}u^{a} &\hspace{.1 in} , \hspace{.1 in} \Theta:=\nabla_{a}u^{a} \hspace{.1 in}, \hspace{.1 in} \omega_{a}:=\tfrac{1}{2}\epsilon_{abc}\tilde{\nabla}^{b}u^{c}.
\end{align}
\end{subequations}

\paragraph{The energy-momentum tensor for a perfect fluid:} 
\begin{subequations}\label{eq:EMT}
\begin{align}
\label{eq:EMT:1}
T_{ab}&=\mu u_{a}u_{b}+p h_{ab}
\\
\label{eq:EMT:2}
\mu:=T_{ab}&u^{a}u^{b} \hspace{.1 in} , \hspace{.1 in} p:=\tfrac{1}{3}T_{ab}h^{ab}.
\end{align}
\end{subequations}

\paragraph{The Riemann curvature tensor:} 
\begin{subequations}\label{eq:RCT}
\begin{align}
\label{eq:RCT:1}
R^{ab}{}_{cd}&=4u^{[a}u_{[c}E^{b]}{}_{d]}+4h^{[a}{}_{[c}E^{b]}{}_{d]}+2\epsilon^{abe}u_{[c}H_{d]e}+2\epsilon_{cde}u^{[a}H^{b]e}\nonumber\\
&\qquad +\tfrac{2}{3}(\mu+3p-2\Lambda)u^{[a}u_{[c}h^{b]}{}_{d]}+\tfrac{2}{3}(\mu+\Lambda)h^{[a}{}_{[c}h^{b]}{}_{d]}
\\\nonumber\\
\label{eq:RCT:2}
E_{ab}&:=C_{cdef}h^{c}{}_{a}u^{d}h^{e}{}_{b}u^{f} \hspace{.1 in} , \hspace{.1 in} H_{ab}:=-\tfrac{1}{2}\epsilon_{cdgh}C^{gh}{}_{ef}h^{c}{}_{a}u^{d}h^{e}{}_{b}u^{f}.
\end{align}
\end{subequations}

\paragraph{The vorticity scalar:} 
\begin{subequations}\label{eq:VS}
\begin{align}
\label{eq:VS:1}
\omega^{2}&=\omega_{a}\omega^{a}\geq 0.
\end{align}
\end{subequations}

\subsection{Ricci identities for a shear-free $(\sigma_{ab}=0)$ perfect fluid}

\subsubsection{Time derivative equations:}
\begin{subequations}\label{eq:RI}
\begin{align}
\label{eq:RI:1}
\dot{\Theta} &= \tilde{\nabla}_{a}\dot{u}^{a} + \dot{u}_{a} \
\dot{u}^{a} -  \tfrac{1}{3} \Theta^2 + 2 \omega^2- \tfrac{1}{2} \bigl(\mu + 3 p\bigr) +\Lambda
\\
\label{eq:RI:2}
\dot{\omega}^{\langle a\rangle} &= \tfrac{1}{2} \epsilon^{abc} \tilde{\nabla}_{b}\dot{u}_{c} - \tfrac{2}{3} \Theta \omega^{a}
\end{align}
\end{subequations}

\subsubsection{Constraint equations:}
\begin{subequations}\label{eq:RICE}
\begin{align}
\label{eq:RICE:1}
0&=(C_1)^{a} := - \tfrac{2}{3} \tilde{\nabla}^{a}\Theta + \epsilon^{a}{}_{bc} \left(\tilde{\nabla}^{b}\omega^{c}+2 \dot{u}^{b} \omega^{c}\right)
\\
\label{eq:RICE:2}
0&=(C_2) :=  \tilde{\nabla}_{a}\omega^{a}- \dot{u}_{a} \omega^{a} 
\\
\label{eq:RICE:3}
0&=(C_3)^{ab} := H^{ab} + 2\dot{u}^{\langle a} \omega^{b\rangle} + \tilde{\nabla}^{\langle a}\omega^{b\rangle}
\end{align}
The new constraint equation arising from setting the shear to zero in equation \eqref{eq:RIG:2} is given as follows,
\begin{equation}
\label{eq:RICE:4}
0=(C_6)^{ab} :=  \tilde{\nabla}^{\langle a}\dot{u}^{b\rangle}+ \dot{u}^{\langle a} \dot{u}^{b\rangle} -  \omega^{\langle a} \omega^{b\rangle} - E^{ab}.
\end{equation}
\end{subequations}
\subsection{(Contracted) second Bianchi identities $(\sigma_{ab}=0)$}
\subsubsection{Time derivative equations:}
\begin{subequations}\label{eq:BI}
\begin{align}
\label{eq:BI:1}
\dot{E}^{\langle ab\rangle} - \epsilon^{cd\langle a} \tilde{\nabla}_{c}H^{b\rangle}{}_{d} &= -  \Theta E^{ab}  + \epsilon^{cd\langle a}\left( 2\dot{u}_{c} H^{b\rangle}{}_{d}  +  \omega_{c}E^{b\rangle}{}_{d}\right)
\\
\label{eq:BI:2}
\dot{H}^{\langle ab\rangle} + \epsilon^{cd\langle a} \tilde{\nabla}_{c}E^{b\rangle}{}_{d} &= -  \Theta H^{ab}  - \epsilon^{cd\langle a}\left( 2\dot{u}_{c} E^{b\rangle}{}_{d}  -  \omega_{c}H^{b\rangle}{}_{d}\right)
\\
\label{eq:BI:3}
\tilde{\nabla}^{a}\mathit{p} &= -\left(\mathit{p} + \mu\right)\dot{u}^{a} 
\\
\label{eq:BI:4}
\dot{\mu} &= -\Theta \left( p +  \mu\right)
\end{align}
\end{subequations}

\subsubsection{Constraint equations:}
\begin{subequations}\label{eq:BICE}
\begin{align}
\label{eq:BICE:1}
0&=(C_4)^{a} := \tilde{\nabla}_{b}E^{ab} - \tfrac{1}{3} \tilde{\nabla}^{a}\mu - 3 H^{a}{}_{b} \omega^{b}
\\
\label{eq:BICE:2}
0&=(C_5)^{a} := \tilde{\nabla}_{b}H^{ab} + (p  + \mu )\omega^{a} + 3 E^{a}{}_{b} \omega^{b}
\end{align}
\end{subequations}
\section{Propagation of the new constraints along $\mathbf u$}\label{Sec5}
The propagation of constraint $(C_{1})^{a}$ along $\bf u$ can be shown to yield equation \eqref{eq:PropCE:1} below by using the following relations:\eqref{TSDCofS},\eqref{TSDCofOmega},\eqref{PropAcc},\eqref{OmegaDot1} and \eqref{eq:RI:1}, this indicates that if constraint $(C_{1})^{a}$ is zero initially, it remains zero at a later time if and only if the $\tilde{\nabla}_{c}(C_6)^{ac}$ vanishes,
\begin{align}
(\dot{C_1})^{\langle a\rangle} + \tilde{\nabla}_{c}(C_6)^{ac}
 &= - \Theta(C_1)^{a} -  \tfrac{1}{2}\epsilon^{a}{}_{cb}(C_1)^{c}\omega^{b} -  \tfrac{8}{3} (C_2) \omega^{a} - 3 (C_3)^{a}{}_{c} \omega^{c} \nonumber\\
 &\qquad - (C_4)^{a} -  (C_6)^{a}{}_{c} \dot{u}^{c}. \label{eq:PropCE:1}
\end{align}
When the scalar constraint equation $(C_{2})$ is propagated along $\bf u$, we obtain equation \eqref{eq:PropCE:2} below, where we have used relations: \eqref{TSDCofOmega}, \eqref{PropAcc} and \eqref{OmegaDot1}. Clearly when $(C_{2})$ is initially zero it remains zero as the universe expands,
\begin{align}
(\dot{C}_ 2) &= - (C_ 2) \Theta. \label{eq:PropCE:2}
\end{align}
Time propagating $(C_3)^{ab}$ and employing the following relations: \eqref{eq:BI:2},\eqref{TSDCofOmega}, \eqref{PropAcc} and \eqref{OmegaDot1}, we arrive at the propagation equation \eqref{eq:PropCE:3} below. Which states that as the universe evolves in time, if $(C_3)^{ab}$ started out being zero it will remain nil at a later time if and only if $\epsilon^{cd\langle a} \tilde{\nabla}_{c}(C_6)^{b\rangle}{}_{d}$ vanishes,
\begin{align}
(\dot{C}_3)^{\langle ab\rangle} - \epsilon^{cd\langle a} \tilde{\nabla}_{c}(C_6)^{b\rangle}{}_{d} &= - \Theta(C_3)^{ab} + \epsilon^{cd\langle a}{} \omega_{c}(C_3)^{b\rangle}{}_{d} + \epsilon^{cd\langle a}{} \dot{u}_{c}(C_6)^{b\rangle}{}_{d}\nonumber\\
&\qquad + \tfrac{3}{2} (C_1)^{\langle a} \omega^{b\rangle}. \label{eq:PropCE:3}
\end{align}
Propagating the constraint equation $(C_4)^{a}$ along $\bf u$ and making use of the following relations: \eqref{TSDCofDivE},\eqref{TSDCofS},\eqref{eq:BI:4},\eqref{eq:BI:2} and \eqref{OmegaDot1}, and applying the identities \eqref{Ident2C4} to \eqref{Ident8C4} this results in equation \eqref{eq:PropCE:4} which shows that $\epsilon^{a}{}_{bc} \tilde{\nabla}^{b}(C_5)^{c}$ must be zero in order for $(C_4)^{a}$ to remain nil with time if it were zero to begin with,
\begin{align}
(\dot{C_4})^{\langle a\rangle} - \tfrac{1}{2} \epsilon^{a}{}_{bc} \tilde{\nabla}^{b}(C_5)^{c} &= -2 \epsilon^{a}{}_{cd}E_{b}{}^{d}(C_3)^{bc} + \epsilon^{a}{}_{cd} H_{b}{}^{d}(C_6)^{bc} -  \tfrac{3}{2} \epsilon^{a}{}_{bc}(C_5)^{b} \dot{u}^{c} \nonumber\\
&\qquad -  \tfrac{4}{3} \Theta (C_4)^{a} - \tfrac{1}{2} \bigl( p + \mu\bigr)(C_1)^{a} + \tfrac{1}{2}\epsilon^{a}{}_{bc}(C_4)^{b}\omega^{c}. \label{eq:PropCE:4}
\end{align}
Complementary, propagating the constraint equation $(C_5)^{a}$ along $\bf u$ and making use of the following relations: \eqref{TSDCofDivH},\eqref{eq:BI:4},\eqref{eq:BI:1} and \eqref{OmegaDot1}, and employing the identities \eqref{ident1C5} to \eqref{ident11C5} this results in equation \eqref{eq:PropCE:5} which shows that $\epsilon^{a}{}_{bc} \tilde{\nabla}^{b}(C_4)^{c}$ must be zero in order for $(C_5)^{a}$ to remain zero as the universe expands,
\begin{align}
(\dot{C_5})^{\langle a\rangle} +  \tfrac{1}{2} \epsilon^{a}{}_{bc} \tilde{\nabla}^{b}(C_4)^{c} &= - \epsilon^{a}{}_{cd}E_{b}{}^{d}(C_6)^{bc} - 2 \epsilon^{a}{}_{cd} H_{b}{}^{d}(C_3)^{bc} + \tfrac{3}{2}\epsilon^{a}{}_{bc}(C_4)^{b} \dot{u}^{c} \nonumber\\
&\qquad -  \tfrac{4}{3}\Theta (C_5)^{a} + \tfrac{1}{2} \epsilon^{a}{}_{bc}(C_5)^{b} \omega^{c}. \label{eq:PropCE:5}
\end{align}
Finally, time propagating the new constraint equation $(C_6)^{ab}$ along $\bf u$ and utilizing the following relations: \eqref{TSDCofAcc},\eqref{eq:BI:1},\eqref{OmegaDot1} and \eqref{PropAcc} to write the resulting terms in terms of the other constraints as much as possible, we arrive at equation \eqref{C6Dot} below. This constraint will form the corner stone of our proof.
\begin{align}
(\dot{C}_6)^{\langle ab\rangle} + \epsilon^{cd\langle a} \tilde{\nabla}_{c}(C_3)^{b\rangle}{}_{d} &= \tfrac{1}{9} \Theta \left(2 - 9 \phi + 3 \psi\right)\dot{u}^{\langle a} \dot{u}^{b\rangle} - \tfrac{4}{3} \Theta \left(1 - 3 \mathit{p}'\right) \omega^{\langle a} \omega^{b\rangle}  \nonumber \\ 
&\qquad + \tfrac{2}{3}\left(1 + 3 \mathit{p}' - 3 \phi\right) \dot{u}^{\langle a}\tilde{\nabla}^{b\rangle}\Theta + \tfrac{1}{3} \Theta \left(2 - 3 \phi\right) \tilde{\nabla}^{\langle a}\dot{u}^{b\rangle} \nonumber\\
&\qquad + \tfrac{1}{3} \left(1 + 3 \mathit{p}'\right) \tilde{\nabla}^{\langle a}\tilde{\nabla}^{b\rangle}\Theta + (C_1)^{\langle a} \dot{u}^{b\rangle} + \tfrac{1}{2}\tilde{\nabla}^{\langle a}(C_1)^{b\rangle} \nonumber\\
&\qquad - 2 \epsilon^{cd\langle a} \dot{u}_{c}(C_3)^{b\rangle}{}_{d} + 2 \epsilon^{cd\langle a} \omega_{c}(C_6)^{b\rangle}{}_{d} - \Theta(C_6)^{ab}.
\label{C6Dot}
\end{align}
From the right hand side of equation \eqref{C6Dot} above, we can extract the non-zero terms which then forms our new constraint $(C_7)^{ab}$ below, 
\begin{align}
0=(C_7)^{ab} &:= \tfrac{1}{9} \Theta \left(2 - 9 \phi + 3 \psi\right)\dot{u}^{\langle a} \dot{u}^{b\rangle} - \tfrac{4}{3} \Theta \left(1 - 3 \mathit{p}'\right) \omega^{\langle a} \omega^{b\rangle} + \tfrac{1}{3} \Theta \left(2 - 3 \phi\right) \tilde{\nabla}^{\langle a}\dot{u}^{b\rangle}  \nonumber \\ 
&\qquad  + \tfrac{1}{3} \left(1 + 3 \mathit{p}'\right) \tilde{\nabla}^{\langle a}\tilde{\nabla}^{b\rangle}\Theta  + \tfrac{2}{3}\left(1 + 3 \mathit{p}' - 3 \phi\right) \dot{u}^{\langle a}\tilde{\nabla}^{b\rangle}\Theta. \label{C6Dot1}
\end{align}
The above equation \eqref{C6Dot1} has been simplified by defining $p^{\prime}=\frac{\partial p}{\partial\mu}$ and using the following scalar variables, $\mathcal{E}$ is the sum of the pressure and the energy density of the fluid, $\mathcal{J}$ is the divergence of the acceleration vector, $\phi$ is related to the second partial derivative of the pressure with respect to the energy density and $\psi$ is related to the third partial derivative of the pressure with respect to the energy density as follows, 
\begin{equation}
\mathcal{E}=\mathit{p}+\mu ,
\label{eq:Energy}
\end{equation}
\begin{equation}
\mathcal{J} = \dot{u}_{b} \dot{u}^{b} + \tilde{\nabla}_{b}\dot{u}^{b} ,
\label{eq:DivAcc}
\end{equation}
\begin{equation}
\phi = \tfrac{1}{3} + \mathcal{E} \tfrac{\mathit{p}''}{\mathit{p}'} -  \mathit{p}' ,
\label{eq:phi}
\end{equation}
\begin{equation}
\psi = -3 \mathit{p}' + 3\mathcal{E}^2 \tfrac{\mathit{p}^{(3)} }{\mathit{p}'^2} + \tfrac{1}{3} (19 - 36 \phi) -  \tfrac{1}{3\mathit{p}'}(4 - 3 \phi) (1 - 3 \phi).
\label{eq:psi}
\end{equation}

\section{The Theorem}\label{Sec6}
\subsection{The case where $p=constant$ with a non-zero cosmological constant $\Lambda$}
\textbf{Theorem 1:} {\it In general relativity, if the velocity vector field of a {\bf geodesic} barotropic perfect fluid is shear-free, then either the expansion or the rotation vanishes.}

\begin{equation}
\nabla_{a}u_{b} = \tfrac{1}{3}\Theta h_{ab} + \epsilon_{abc} \omega^{c}
\end{equation}
\begin{equation}
\mu + p\neq 0,\hspace{.1 in} p=constant,\hspace{.1 in} \dot{u}^{a}=0 \hspace{.1 in}\text{and} \hspace{.1 in} \sigma_{ab}=0 \hspace{.25 in} \Longrightarrow \hspace{.25 in}\omega\theta=0.
\end{equation}
{\it Proof:}
For this proof we shall focus on the time propagation of constraint $(C_{1})^{a}$ which was provided in reference \cite{senovilla1998theorems}. From equation \eqref{eq:PropCE:1} we can see that the new constraint comes from the divergence of $(C_{6})^{ab}$ as follows;
\begin{align}
\tilde{\nabla}_{c}(C_6)^{ac} &= - (C_4)^{a} - 3 (C_3)^{a}{}_{c} \omega^{c} - 2 (C_2) \omega^{a} + \tfrac{1}{2} \epsilon^{a}{}_{cb}(C_1)^{c} \omega^{b}  -  \tfrac{1}{3} \tilde{\nabla}^{a}\mu + \tfrac{8}{3} \omega \tilde{\nabla}^{a}\omega  \nonumber\\
&\qquad -  \tfrac{1}{3} \epsilon^{a}{}_{cb} \omega^{c} \tilde{\nabla}^{b}\Theta. \label{DivC6}
\end{align}
From equation \eqref{DivC6}, the non-zero terms on the right hand side form a new constraint equation shown below,
\begin{equation}
0=(C_{\text{v7}})^{a} :=  \tilde{\nabla}^{a}\mu - 8\omega \tilde{\nabla}^{a}\omega + \epsilon^{a}{}_{cb} \omega^{c} \tilde{\nabla}^{b}\Theta. \label{Cv7}
\end{equation}
Time propagating equation \eqref{Cv7} and applying relation: \eqref{TSDCofS}, \eqref{TSDCofOmega}, \eqref{TSDCofS}, \\
\eqref{OmegaDot1} and \eqref{OmegaDot2}, and applying \eqref{Cv7} itself, we obtain the follow,
\begin{align}
(\dot{C}_{\text{v7}})^{a} &= - \tfrac{5}{3} (C_{\text{v7}})^{a} \Theta -  \tfrac{1}{2} \epsilon^{a}{}_{cb}(C_{\text{v7}})^{c} \omega^{b} - \left( \mathit{p} +  \mu - \tfrac{29}{6} \omega^2\right) \tilde{\nabla}^{a}\Theta + \tfrac{1}{3} \Theta \tilde{\nabla}^{a}\mu \nonumber\\
&\qquad + \tfrac{1}{2} \omega^{a} \omega^{c} \tilde{\nabla}_{c}\Theta. \label{Cv7Dot}
\end{align}
In equation \eqref{Cv7Dot}, the terms that can not be expressed in terms of the other constraints, result in a new constraint equation below,
\begin{equation}
0=(C_{\text{v8}})^{a} := \left( \mathit{p} +  \mu - \tfrac{29}{6} \omega^2\right) \tilde{\nabla}^{a}\Theta - \tfrac{1}{3} \Theta \tilde{\nabla}^{a}\mu - \tfrac{1}{2} \omega^{a} \omega^{c} \tilde{\nabla}_{c}\Theta. \label{Cv8}
\end{equation}
Moving on, we compute the time propagation of \eqref{Cv8} with the aid of the following relations: \eqref{TSDCofS},\eqref{TSDCofS},\eqref{OmegaDot1},\eqref{OmegaDot2},\eqref{Cv7},\eqref{eq:BI:4} and \eqref{Cv8} itself we arrive at,
\begin{align}
12\Theta(\dot{C}_{\text{v8}})^{a} &= -2(C_{\text{v8}})^{a} \big(9 \mathit{p} - 6 \Lambda + 10 \Theta^2 + 3 \mu - 12 \omega^2\big) - (C_{\text{v7}})^{a} \Theta \big(6 \mathit{p} + 6 \mu - 29 \omega^2\big) \nonumber \\ 
&\qquad - 12 \epsilon^{a}{}_{cb}(C_{\text{v8}})^{c} \Theta \omega^{b} + 3 (C_{\text{v7}})^{c} \Theta \omega^{a} \omega_{c} + \big(18 \mathit{p}^2 - 12 \mathit{p} \Lambda + 24 \mathit{p} \mu - 12 \Lambda \mu \nonumber \\ 
&\qquad + 6 \mu^2 - 111 \mathit{p} \omega^2 + 58 \Lambda \omega^2 + \tfrac{116}{3} \Theta^2 \omega^2 - 53 \mu \omega^2 + 116 \omega^4\big) \tilde{\nabla}^{a}\Theta  \nonumber \\ 
&\qquad + \Theta\epsilon^{a}{}_{cb} \omega^{c} \tilde{\nabla}^{b}\Theta \big(6 \mathit{p} + 6 \mu - 29 \omega^2\big)  - \big(9 \mathit{p} - 6 \Lambda - 4 \Theta^2 + 3 \mu \nonumber\\
&\qquad - 12 \omega^2\big) \omega^{a} \omega^{c} \tilde{\nabla}_{c}\Theta .\label{Cv8Dot}
\end{align}
In equation \eqref{Cv8Dot}, we can extract the non-zero terms to form a new consistency constraint below,
 \begin{align}
0=(C_{\text{v9}})^{a} &:= \big(18 \mathit{p}^2 - 12 \mathit{p} \Lambda + 24 \mathit{p} \mu - 12 \Lambda \mu + 6 \mu^2 - 111 \mathit{p} \omega^2 + 58 \Lambda \omega^2 + \tfrac{116}{3} \Theta^2 \omega^2 \nonumber \\ 
&\qquad  - 53 \mu \omega^2 + 116 \omega^4\big) \tilde{\nabla}^{a}\Theta + \Theta\epsilon^{a}{}_{cb} \omega^{c} \tilde{\nabla}^{b}\Theta \big(6 \mathit{p} + 6 \mu - 29 \omega^2\big)  \nonumber\\
&\qquad - \big(9 \mathit{p} - 6 \Lambda - 4 \Theta^2 + 3 \mu - 12 \omega^2\big) \omega^{a} \omega^{c} \tilde{\nabla}_{c}\Theta .\label{Cv9}
\end{align}
In order to reduce the number of terms in equation \eqref{Cv9} above, we make the observation that the second term which contains $\epsilon^{a}{}_{cb} \omega^{c} \tilde{\nabla}^{b}\Theta$ can be eliminated by contracting with $\omega^{a}$, since they are orthogonal to each other. This action leads us to the following scalar constraint equation,
\begin{align}
0=(C_{\text{s9}}) &:= \big(\tfrac{9}{2} \mathit{p}^2 - 3 \mathit{p} \Lambda + 6 \mathit{p} \mu - 3 \Lambda \mu + \tfrac{3}{2} \mu^2 - 30 \mathit{p} \omega^2 + 16 \Lambda \omega^2 + \tfrac{32}{3} \Theta^2 \omega^2 \nonumber \\ 
&\qquad - 14 \mu \omega^2 + 32 \omega^4\big) \omega^{a} \tilde{\nabla}_{a}\Theta .\label{Cs9}
\end{align}
Therefore we have either that $\omega^{a} \tilde{\nabla}_{a}\Theta=0$ or the term in the parenthesis vanishes in equation \eqref{Cs9}. We shall consider these cases separately. 
When the terms within the brackets in equation \eqref{Cs9} vanish, this gives us the following constraint,
\begin{align}
0=(C_{\text{s10}}) &:= \tfrac{9}{2} \mathit{p}^2 - 3 \mathit{p} \Lambda + 6 \mathit{p} \mu - 3 \Lambda \mu + \tfrac{3}{2} \mu^2 - 30 \mathit{p} \omega^2 + 16 \Lambda \omega^2 + \tfrac{32}{3} \Theta^2 \omega^2 \nonumber \\ 
&\qquad - 14 \mu \omega^2 + 32 \omega^4 .\label{Cs10}
\end{align}
Time propagating equation \eqref{Cs10} and making use of the following relations \eqref{OmegaDot2}, \eqref{eq:BI:4}, \eqref{eq:RI:1} and \eqref{Cs9} we obtain the following,
\begin{align}
(\dot{C}_{\text{s10}}) &= \tfrac{1}{3} \Theta \big(9 \mathit{p}^2 - 9 \mathit{p} \Lambda - 6 (C_{\text{s10}}) + 9 \mathit{p} \mu - 9 \Lambda \mu - 114 \mathit{p} \omega^2 + 96 \Lambda \omega^2 - 18 \mu \omega^2 \nonumber \\ 
&\qquad + 64 \omega^4\big). \label{Cs10Dot}
\end{align}
From equation \eqref{Cs10Dot} above, we can extract the non-zero terms from the right hand side to form a new constraint given by,
\begin{align}
0=(C_{\text{s11}}) &:= \Theta \big(9 \mathit{p}^2 - 9 \mathit{p} \Lambda + 9 \mathit{p} \mu - 9 \Lambda \mu - 114 \mathit{p} \omega^2 + 96 \Lambda \omega^2 - 18 \mu \omega^2 + 64 \omega^4\big), \label{Cs11}
\end{align}
where we have that $\Theta=0$ and we have the proof, or that the terms in the brackets must be zero. Then from the latter we form the following consistency constraint equation, 
\begin{align}
0=(C_{\text{s12}}) &:= 9\big(\mathit{p}^2 - \mathit{p} \Lambda  + \mathit{p} \mu - \Lambda \mu\big) - 2\big(57 \mathit{p} - 48 \Lambda + 9 \mu - 32 \omega^2\big)\omega^2. \label{Cs12}
\end{align}
Further time propagating equation \eqref{Cs12}, and making use of relations \eqref{eq:BI:4}, \eqref{OmegaDot2} and \eqref{Cs12} itself gives us the following,
\begin{equation}
\tfrac{1}{24} (\dot{C}_{\text{s12}}) = \Theta \Big(- \tfrac{1}{24} (C_{\text{s12}}) + \big( \mu + \tfrac{7}{3} \mathit{p}  -  \tfrac{40}{9} \omega^2 - \tfrac{4}{3} \Lambda \big)\omega^2\Big) \label{Cs12Dot}
\end{equation}
We can extract the non-zero terms from equation \eqref{Cs12Dot} above to obtain a constraint equation similar to equation (48) in \cite{senovilla1998theorems}, given below as, 
\begin{equation}
0=(C_{\text{s13}}) := \Theta \big( \mu + \tfrac{7}{3} \mathit{p}  -  \tfrac{40}{9} \omega^2 - \tfrac{4}{3} \Lambda \big)\omega^2. \label{Cs13}
\end{equation}
From equation \eqref{Cs13} we observe that for consistency we have that $\Theta\omega=0$ in which we have the proof, or the terms in the brackets must vanish. In what follows we shall consider the latter as our new constraint,
\begin{equation}
0=(C_{\text{s14}}) :=  \mu + \tfrac{7}{3} \mathit{p}  -  \tfrac{40}{9} \omega^2 - \tfrac{4}{3} \Lambda. \label{Cs14}
\end{equation} 
The computation of the time propagation of equation \eqref{Cs14}, with the aid of equation \eqref{eq:BI:4} and \eqref{OmegaDot2} gives us the following,
\begin{equation}
(\dot{C}_{\text{s14}}) = -\Theta \big(\mathit{p} +  \mu - \tfrac{160}{27} \omega^2\big) . \label{Cs14Dot}
\end{equation}
In equation \eqref{Cs14Dot} we require that the terms in the parenthesis on the right hand side be zero, which then becomes our new constraint equation as folllows,
\begin{equation}
0=(C_{\text{s15}}) := \mathit{p} +  \mu - \tfrac{160}{27} \omega^2 .\label{Cs15}
\end{equation} 
Time propagating equation \eqref{Cs15}, and using relations \eqref{eq:BI:4}, \eqref{OmegaDot2} and \eqref{Cs15} itself one obtains,
\begin{equation}
(\dot{C}_{\text{s15}}) = - (C_{\text{s15}}) \Theta + \tfrac{160}{81} \Theta \omega^2. \label{Cs15Dot}
\end{equation}
From equation \eqref{Cs15Dot}, we arrive at the following restriction for $\Theta$ and  $\omega$,  
\begin{equation}
0=(C_{\text{s16}}) :=  \tfrac{160}{81} \Theta \omega^2. \label{Cs16}
\end{equation}
Hence $\Theta \omega=0$, and we have the proof. For completion we need to also consider the other half of equation \eqref{Cs9}, where if the terms in the brackets  do not vanish, then it is required that $\omega^{a}\tilde{\nabla}_{a}\Theta=0$. Therefore, we can see from equation \eqref{Cv9} that this condition leads to the following constraint,
 \begin{align}
0=(C_{\text{v10}})^{a} &:= \big(18 \mathit{p}^2 - 12 \mathit{p} \Lambda + 24 \mathit{p} \mu - 12 \Lambda \mu + 6 \mu^2 - 111 \mathit{p} \omega^2 + 58 \Lambda \omega^2 + \tfrac{116}{3} \Theta^2 \omega^2 \nonumber \\ 
&\qquad  - 53 \mu \omega^2 + 116 \omega^4\big) \tilde{\nabla}^{a}\Theta + \Theta\epsilon^{a}{}_{cb} \omega^{c} \tilde{\nabla}^{b}\Theta \big(6 \mathit{p} + 6 \mu - 29 \omega^2\big).\label{Cv10}
\end{align}
Since $\tilde{\nabla}^{a}\Theta$ and $\epsilon^{a}{}_{cb} \omega^{c} \tilde{\nabla}^{b}\Theta$ are orthogonal vectors, contracting equation \eqref{Cv10} with $\epsilon_{acb} \omega^{c} \tilde{\nabla}^{b}\Theta$ and applying the condition that $\omega^{a}\tilde{\nabla}_{a}\Theta=0$ yields the following constraint equation,
\begin{equation}
0=(C_{\text{s17}}) := \tfrac{3}{2} \Theta \omega^2 \big( \mathit{p} +  \mu - \tfrac{29}{6} \omega^2\big) \tilde{\nabla}_{a}\Theta \tilde{\nabla}^{a}\Theta. \label{Cs17}
\end{equation}
Equation \eqref{Cs17} presents us with three cases. Either (1) $\Theta\omega=0$, which proves the theorem, or (2) the term in parenthesis must vanish, which forms a  constraint
\begin{equation}
0=(C_{\text{s18}}) := \mathit{p} +  \mu - \tfrac{29}{6} \omega^2, \label{Cs18}
\end{equation}
or (3) $\tilde{\nabla}_{a}\Theta \tilde{\nabla}^{a}\Theta=0$. In case (2) we need to time propagate equation \eqref{Cs18} once more in order to reach the restriction on the expansion and vorticity scalars, this gives us 
\begin{equation}
(\dot{C}_{\text{s18}}) = - (C_{\text{s18}}) \Theta + \tfrac{29}{18} \Theta \omega^2 \label{Cs18Dot}
\end{equation}
where just like in equation \eqref{Cs15Dot} above, we have that $\Theta \omega=0$, which completes the proof. Finally, in case (3) we can see that the spatial gradient vector of the expansion scalar is orthogonal to itself, hence $\tilde{\nabla}^{a}\Theta=0$, then by equation \eqref{Cv7} and \eqref{Cv8} it follows that $\tilde{\nabla}^{a}\mu=\tilde{\nabla}^{a}\omega=0$. This leads according to the relation \eqref{SSCofS} to $\dot{\omega}=0$, which implies that the vorticity is constant and that $\Theta \omega=0$ by equation \eqref{OmegaDot2}. Alternatively, consider taking the spatial divergence of constraint $(C_{1})^{a}$ and using the relation \eqref{SSDCofOmega} this yields the following;
\begin{equation}
0=\tilde{\nabla}_{a}(C_1)^{a} := - \tfrac{2}{3} \Theta \omega^2 -  \tfrac{2}{3} \tilde{\nabla}_{a}\tilde{\nabla}^{a}\Theta
\end{equation}
where it is clear that when the spatial gradient of the expansion is zero, then $\Theta\omega$ must vanish. This completes the proof. 
\subsection{The conjecture}
\textbf{Conjecture:} {\it In general relativity, if the velocity vector field of a barotropic perfect fluid is shear-free, then either the expansion or the rotation vanishes.}

\begin{equation}
\nabla_{a}u_{b} = - u_{a} \dot{u}_{b} + \tfrac{1}{3}\Theta h_{ab} + \epsilon_{abc} \omega^{c}
\end{equation}

\begin{equation}
\mu + p\neq 0,\hspace{.1 in} p=p(\mu) \hspace{.1 in}\text{and} \hspace{.1 in} \sigma_{ab}=0 \hspace{.25 in} \Longrightarrow \hspace{.25 in}\omega\theta=0.
\end{equation}

\subsubsection{Time propagation of constraint $(C_{6})^{ab}$ and it's spatial gradients:}
From the time propagation of equation \eqref{eq:BI:3} and making use of the relations \eqref{TSDCofS}, \eqref{eq:BI:4} and \eqref{eq:BI:3} itself, we obtain the following propagation equation for the acceleration vector,
\begin{equation}
\ddot{u}^{\langle a\rangle} = - \dot{u}^{a} \Theta \phi -  \epsilon^{a}{}_{bc} \dot{u}^{b} \omega^{c} + \mathit{p}' \tilde{\nabla}^{a}\Theta.
\label{PropAcc}
\end{equation}
We can compute the time evolution of $\mathcal{J}$ in equation \eqref{eq:DivAcc}, by using equation \eqref{PropAcc}, this given the following,
\begin{align}
\dot{\mathcal{J}} &= - \mathcal{J} \Theta ( \tfrac{1}{3} +  \phi) + \tfrac{1}{9} \dot{u}_{a} \dot{u}^{a} \Theta (5 + 3 \psi) - \mathit{p}' (1 - 9 \mathit{p}') \Theta \omega^2 \nonumber \\ 
&\qquad + (1 - 2 \phi) \dot{u}^{a}\tilde{\nabla}_{a}\Theta-3 \mathit{p}'(C_1)^{a}\dot{u}_{a}  -  \tfrac{3}{2} \mathit{p}' \tilde{\nabla}_{a}(C_1)^{a}. \label{JDot}
\end{align}
When substituting for $\dot{u}^a$ from equation \eqref{eq:BI:3} in equation \eqref{eq:RI:2} and applying equation \eqref{SSCofS}, \eqref{eq:BI:4} and \eqref{eq:BI:3} we obtain the following evolution equation for the vorticity vector and scalar respectively,
\begin{equation}
\dot{\omega}^{\langle a\rangle} =  \Theta\omega^{a}\bigl( \mathit{p}'-\tfrac{2}{3}\bigr) \label{OmegaDot1}
\end{equation}
\begin{equation}
\dot{\omega} =  \Theta\omega\bigl( \mathit{p}'-\tfrac{2}{3}\bigr). \label{OmegaDot2} 
\end{equation}
Using the time propagation of constraint $(C_6)^{ab}$ along $\bold{u}$ given by equation \eqref{C6Dot}, and extracting the non-zero terms we can form a new constraint $(C_7)^{ab}$ which is given as follows,
\begin{align}
0=(C_7)^{ab} &:= \tfrac{1}{9} \Theta \left(2 - 9 \phi + 3 \psi\right)\dot{u}^{\langle a} \dot{u}^{b\rangle} - \tfrac{4}{3} \Theta \left(1 - 3 \mathit{p}'\right) \omega^{\langle a} \omega^{b\rangle} + \tfrac{1}{3} \Theta \left(2 - 3 \phi\right) \tilde{\nabla}^{\langle a}\dot{u}^{b\rangle}  \nonumber \\ 
&\qquad  + \tfrac{1}{3} \left(1 + 3 \mathit{p}'\right) \tilde{\nabla}^{\langle a}\tilde{\nabla}^{b\rangle}\Theta  + \tfrac{2}{3}\left(1 + 3 \mathit{p}' - 3 \phi\right) \dot{u}^{\langle a}\tilde{\nabla}^{b\rangle}\Theta .
\label{C7ab}
\end{align}
Equation \eqref{C7ab} contains terms involving the first and second spatial gradients of the expansion scalar together with the acceleration vector and it's spatial gradient, these terms are problematic under time propagation because of \eqref{eq:RI:1}, this shall be discussed later. Therefore, we would like to eliminate these terms if possible.\\
\newline
Time propagating $(C_1)^{a}$ in equation \eqref{eq:PropCE:1} and making use of equations \eqref{SSDCofAcc}, \eqref{eq:BICE:1},\eqref{eq:RICE:3} and \eqref{eq:RICE:4} itself yields the divergence of $(C_6)^{ab}$ as follows,
\begin{align}
\tilde{\nabla}_{b}(C_6)^{ab} &= \tfrac{2}{9}\left(3 \Lambda + 3 \mathcal{J} -  \Theta^2 + 3 \mu + 6 \omega^2\right)\dot{u}^{a}  - \tfrac{1}{3} \Theta \left(2 - 9 \phi\right) \epsilon^{a}{}_{bc} \dot{u}^{b}\omega^{c} \nonumber\\ 
&\qquad + \tfrac{1}{3} \left(2 \tilde{\nabla}^{a}\mathcal{J} - 2 \mathit{p}' \Theta \tilde{\nabla}^{a}\Theta - \tilde{\nabla}^{a}\mu + 8 \omega \tilde{\nabla}^{a}\omega\right) - (C_4)^{a} \nonumber\\
&\qquad - \tfrac{1}{3}\left(1 - 9 \mathit{p}'\right) \epsilon^{a}{}_{bc}\omega^{b} \tilde{\nabla}^{c}\Theta -  (C_6)^{a}{}_{b} \dot{u}^{b} - \mathit{p}' \Theta  (C_1)^{a} \nonumber \\ 
&\qquad  - 2 (C_2) \omega^{a} - 3 (C_3)^{a}{}_{b} \omega^{b} + \tfrac{1}{2}\epsilon^{a}{}_{bc}(C_1)^{b}\omega^{c}.
\label{DivC6ab}
\end{align}
Now from equation \eqref{DivC6ab} above, we can extract the non-zero terms which form the following new constraint equation, 
\begin{align}
0=(C_ 8)^{a} &:= \tfrac{2}{9} \left(3 \Lambda + 3 \mathcal{J} -  \Theta^2 + 3 \mu + 6 \omega^2\right)\dot{u}^{a}  - \tfrac{1}{3} \Theta \left(2 - 9 \phi\right) \epsilon^{a}{}_{bc} \dot{u}^{b}\omega^{c} \nonumber \\ 
&\qquad + \tfrac{1}{3} \left(2 \tilde{\nabla}^{a}\mathcal{J} - 2 \mathit{p}' \Theta \tilde{\nabla}^{a}\Theta -  \tilde{\nabla}^{a}\mu + 8 \omega \tilde{\nabla}^{a}\omega\right)  \nonumber \\ 
&\qquad - \tfrac{1}{3} \left(1 - 9 \mathit{p}'\right) \epsilon^{a}{}_{bc}\omega^{b} \tilde{\nabla}^{c}\Theta. \label{C8}
\end{align}
On taking the curl of $(C_ 8)^{a}$ equation \eqref{C8} above and making use of the following relations: \eqref{SSCofS}, \eqref{SSCofS},\eqref{SSCofS}\eqref{SSCofS} and \eqref{GradComAcc}, we obtain the following constraint equation,
\begin{align}
\epsilon^{a}{}_{bc} \tilde{\nabla}^{b}(C_{8})^{c} &= \epsilon^{a}{}_{bc}(C_{8})^{b}\dot{u}^{c} + \Bigl[(1 - 13 \mathit{p}')  (C_1)^{b} \dot{u}_{b} + \tfrac{1}{2} (1 - 13 \mathit{p}') \tilde{\nabla}_{b}(C_1)^{b} \Bigr]\omega^{a} \nonumber \\ 
&\qquad  +\tfrac{1}{3} (C_2) \Bigl[(1 - 9 \mathit{p}') \tilde{\nabla}^{a}\Theta - \Theta (2 - 9 \phi)\dot{u}^{a}  \Bigr] - \tfrac{1}{3} \Theta (2 - 9 \phi) \omega^{b} \tilde{\nabla}_{b}\dot{u}^{a}\nonumber \\ 
&\qquad + \tfrac{1}{3} \Bigl[ 2\Theta (1 + 3 \mathit{p}') \mathcal{E} + \tfrac{1}{3} \Theta (2 - 39 \phi)\mathcal{J} - \tfrac{1}{9} \Theta (16 - 162 \phi - 39 \psi)\dot{u}_{b} \dot{u}^{b}\nonumber \\ 
&\qquad - \tfrac{1}{3}\Theta  \left(29 + 18 \mathit{p}' - 351 \mathit{p}'^2\right) \omega^2  + 2(5 - 9 \mathit{p}'- 13 \phi) \dot{u}^{b} \tilde{\nabla}_{b}\Theta\Bigr]\omega^{a} \nonumber \\ 
&\qquad - \Bigl[\Theta \psi \dot{u}^{b} \omega_{b} + \tfrac{1}{3} (2 - 9 \phi) \omega^{b} \tilde{\nabla}_{b}\Theta\Bigr]\dot{u}^{a} + \tfrac{1}{3} \Theta (2 - 9 \phi)  \dot{u}^{b}\tilde{\nabla}_{b}\omega^{a} \nonumber \\ 
&\qquad + \tfrac{2}{9} \Theta (1 + 3 \phi) \epsilon^{a}{}_{bc} \dot{u}^{b}\tilde{\nabla}^{c}\Theta - \tfrac{1}{3} (1 + 9 \mathit{p}'- 9 \phi) \dot{u}^{b} \omega_{b} \tilde{\nabla}^{a}\Theta \nonumber \\ 
&\qquad + \tfrac{1}{3} (1 - 9 \mathit{p}') \omega^{b} \tilde{\nabla}_{b}\tilde{\nabla}^{a}\Theta - \tfrac{1}{3} (1 - 9 \mathit{p}') \tilde{\nabla}^{b}\Theta \tilde{\nabla}_{b}\omega^{a}. \label{CurlC8}
\end{align}
From the non-zero terms in equation \eqref{CurlC8}, we can extract a new constraint equation $(C_{9})^{a}$ below, which is actually the curl of the divergence of $(C_{6})^{ab}$ in equation \eqref{eq:RICE:4},
\begin{align}
0=(C_{9})^{a} &:= \tfrac{1}{3}\Bigl[ 2\Theta (1 + 3 \mathit{p}') \mathcal{E} + \tfrac{1}{3} \Theta (2 - 39 \phi)\mathcal{J} - \tfrac{1}{9} \Theta (16 - 162 \phi - 39 \psi)\dot{u}_{b} \dot{u}^{b}\nonumber\\ 
&\qquad - \tfrac{1}{3}\Theta  \left(29 + 18 \mathit{p}' - 351 \mathit{p}'^2\right) \omega^2  + 2(5 - 9 \mathit{p}'- 13 \phi) \dot{u}^{b} \tilde{\nabla}_{b}\Theta\Bigr]\omega^{a}  \nonumber \\ 
&\qquad - \Bigl[\Theta \psi \dot{u}^{b} \omega_{b} + \tfrac{1}{3} (2 - 9 \phi) \omega^{b} \tilde{\nabla}_{b}\Theta\Bigr]\dot{u}^{a} - \tfrac{1}{3} \Theta (2 - 9 \phi) \omega^{b} \tilde{\nabla}_{b}\dot{u}^{a} \nonumber \\ 
&\qquad + \tfrac{1}{3} \Theta (2 - 9 \phi)  \dot{u}^{b}\tilde{\nabla}_{b}\omega^{a} - \tfrac{1}{3} (1 + 9 \mathit{p}'- 9 \phi) \dot{u}^{b} \omega_{b} \tilde{\nabla}^{a}\Theta \nonumber \\ 
&\qquad + \tfrac{2}{9} \Theta (1 + 3 \phi) \epsilon^{a}{}_{bc} \dot{u}^{b}\tilde{\nabla}^{c}\Theta - \tfrac{1}{3} (1 - 9 \mathit{p}') \tilde{\nabla}^{b}\Theta \tilde{\nabla}_{b}\omega^{a} \nonumber\\
&\qquad + \tfrac{1}{3} (1 - 9 \mathit{p}') \omega^{b} \tilde{\nabla}_{b}\tilde{\nabla}^{a}\Theta. \label{C9}
\end{align}
From the last term on the right hand side of equation \eqref{C9} one can see that we have an expression for the double spatial derivative of the expansion scalar $\Theta$. This brings us one step closer to our goal, which is to try and eliminate these terms in equation \eqref{C7ab}. 
Upon taking the symmetric trace-free curl of constraint $(C_{6})^{ab}$ in equation \eqref{eq:RICE:4}, which is term appearing on the left hand side equation \eqref{eq:PropCE:3}, this results in the following equation,
\begin{align}
\epsilon^{\langle a}{}_{cd} \tilde{\nabla}^{c}(C_6)^{b\rangle d} &= \epsilon^{\langle a}{}_{cd}\dot{u}^{c}(C_6)^{b\rangle d} - (C_1)^{(a} \omega^{b)} - 6 \Theta\mathit{p}'(C_2)h^{ab} - 3 \Theta \phi \dot{u}^{\langle a} \omega^{b\rangle}\nonumber \\ 
&\qquad  +  \mathit{p}' \Theta \tilde{\nabla}^{(a}\omega^{b)} - (\tfrac{2}{3}  - 3 \mathit{p}')\omega^{(a} \tilde{\nabla}^{b)}\Theta + \epsilon^{\langle a}{}_{cd} \tilde{\nabla}^{d}E^{b\rangle c} \nonumber \\ 
&\qquad  - 2 \epsilon^{(a}{}_{cd} \dot{u}^{c} \tilde{\nabla}^{d}\dot{u}^{b)} + \epsilon^{(a}{}_{cd} \omega^{c} \tilde{\nabla}^{d}\omega^{b)}  \nonumber \\ 
&\qquad  + \mathit{p}'\bigl( \Theta \dot{u}^{c}\omega_{c} - 6  \omega^{c} \tilde{\nabla}_{c}\Theta \bigr)h^{ab}. \label{CurlC6} 
\end{align}
In the same spirit as before, we extract the non-zero terms from equation \eqref{CurlC6}, which forms a new constraint $(C_{10})^{ab}$ below,
\begin{align}
0=(C_{10})^{ab} &:= - 3 \Theta \phi \dot{u}^{\langle a} \omega^{b\rangle} +  \mathit{p}' \Theta \tilde{\nabla}^{(a}\omega^{b)} - (\tfrac{2}{3}  - 3 \mathit{p}')\omega^{(a} \tilde{\nabla}^{b)}\Theta + \epsilon^{\langle a}{}_{cd} \tilde{\nabla}^{d}E^{b\rangle c} \nonumber \\ 
&\qquad  - 2 \epsilon^{(a}{}_{cd} \dot{u}^{c} \tilde{\nabla}^{d}\dot{u}^{b)} + \epsilon^{(a}{}_{cd} \omega^{c} \tilde{\nabla}^{d}\omega^{b)} + \mathit{p}'\bigl( \Theta \dot{u}^{c}\omega_{c} - 6  \omega^{c} \tilde{\nabla}_{c}\Theta \bigr)h^{ab}. \label{C10}
\end{align}
Take the divergence of $(C_{10})^{ab}$ equation \eqref{C10}, we can see that the third and the eighth terms gives us second spatial derivatives of $\Theta$, which is what we are looking for. To obtain the following equation we have used the following key relations: \eqref{SSDCofOmega}, \eqref{SSDCofAcc} and \eqref{SSDCofEab} together with the identity \eqref{IdentDivC10},
\begin{align}
\tilde{\nabla}_{b}(C_{10})^{ab} &=  -\tfrac{1}{6} \big(2 - 9 \phi\big) \Theta \dot{u}^{b}\tilde{\nabla}_{b}\omega^{a} - \tfrac{1}{2} \big(2 - \phi\big) \Theta \omega^{b} \tilde{\nabla}_{b}\dot{u}^{a} \nonumber \\ 
&\qquad + \tfrac{1}{6} \big(1 - 9 \mathit{p}'\big) \tilde{\nabla}^{b}\Theta\tilde{\nabla}_{b}\omega^{a}  - \tfrac{1}{6} \big(5 + 3 \mathit{p}'\big) \omega^{b} \tilde{\nabla}_{b}\tilde{\nabla}^{a}\Theta\nonumber \\
&\qquad - \tfrac{1}{2} \big(1  +  \mathit{p}'- \phi \big) \dot{u}^{b} \omega_{b} \tilde{\nabla}^{a}\Theta - \tfrac{1}{9} \big(1 + 3 \phi\big) \Theta\epsilon^{a}{}_{bc}\dot{u}^{b} \tilde{\nabla}^{c}\Theta\nonumber \\
&\qquad + \Big[\tfrac{1}{6} \big(2 + 9 \phi\big) \Theta\mathcal{J} - \tfrac{1}{3}  \big(1 + 3 \mathit{p}'\big) \Theta\mathcal{E} - \tfrac{1}{2} \big(6 \phi +  \psi\big) \Theta\dot{u}_b\dot{u}^b  \nonumber \\ 
&\qquad + \tfrac{1}{6} \big(19 - 18 \mathit{p}' - 81 \mathit{p}'^2\big) \Theta \omega^2 - \tfrac{1}{3} \big(5  - 9 \mathit{p}'- 9 \phi\big) \dot{u}^{b} \tilde{\nabla}_{b}\Theta\Big]\omega^{a}  \nonumber \\ 
&\qquad -  \Big[\tfrac{1}{18} \big(8 - 36 \phi + 3 \psi\big) \dot{u}^{b} \Theta \omega_{b} + \tfrac{1}{6} \big(2  + 12 \mathit{p}'- 3 \phi \big) \omega^{b} \tilde{\nabla}_{b}\Theta\Big]\dot{u}^{a} \nonumber \\  
&\qquad - \epsilon^{a}{}_{bc} \Big[ \omega^{b} \tilde{\nabla}^{c}(C_2) + \tfrac{3}{2} (C_{\text{8}})^{b} \dot{u}^{c} +  \tfrac{1}{2}\mathit{p}' \Theta \tilde{\nabla}^{c}(C_1)^{b} +  \tfrac{1}{2}\tilde{\nabla}^{c}(C_4)^{b}\Big]  \nonumber \\ 
&\qquad  + \Big[2 \epsilon^{a}{}_{cd}(C_3)_{b}{}^{d} -  \epsilon_{bcd}(C_3)^{ad}\Big] \dot{u}^{b} \omega^{c} - 2 \epsilon^{a}{}_{cd}(C_6)_{b}{}^{d} \Big[ \dot{u}^{b} \dot{u}^{c} + \omega^{b} \omega^{c}\Big] \nonumber \\
&\qquad - \tfrac{1}{4} \omega^{b} \Big[2 \tilde{\nabla}^{a}(C_1)_{b} + 5 \tilde{\nabla}_{b}(C_1)^{a} + 2 \epsilon_{bcd} \tilde{\nabla}^{d}(C_3)^{ac} + 12 \epsilon^{a}{}_{cd} \tilde{\nabla}^{d}(C_3)_{b}{}^{c}\Big] \nonumber \\
&\qquad - (C_1)^{b} \Big[ \tfrac{1}{4}\tilde{\nabla}_{b}\omega^{a} +  \tfrac{1}{2} \epsilon^{a}{}_{bc} \mathit{p}' \tilde{\nabla}^{c}\Theta + \tfrac{1}{2} (5 - 9 \mathit{p}') \dot{u}_{b} \omega^{a}  + \tfrac{3}{2}(C_3)^{a}{}_{b} +  \dot{u}^{a} \omega_{b} \nonumber \\ 
&\qquad  - \tfrac{1}{6} \epsilon^{a}{}_{bc} \big(1 + 3 \phi + 9 \mathit{p}'\big) \dot{u}^{c} \Theta\Big] + \epsilon^{a}{}_{bd} \Big[ - \tfrac{1}{2} \omega^{b} \tilde{\nabla}_{c}(C_3)^{cd} + \tfrac{3}{2} (C_3)_{c}{}^{d} \tilde{\nabla}^{c}\omega^{b}\Big] \nonumber \\ 
&\qquad + \tfrac{1}{4} \Big[(C_1)^{a} \dot{u}^{b} \omega_{b} - 4 \mathit{p}' \Theta \tilde{\nabla}^{a}(C_2) - 4 (C_3)^{a}{}_{b} \tilde{\nabla}^{b}\Theta\Big]- \tfrac{1}{4} \big(2 - 9 \mathit{p}'\big) \omega^a\tilde{\nabla}_{b}(C_1)^{b}  \nonumber \\ 
&\qquad + \tfrac{1}{2}(C_2) \Big[\big(3 \phi - 2 \mathit{p}'\big) \dot{u}^{a} \Theta + \tfrac{1}{2}(C_1)^{a} + \big(1 - 3 \mathit{p}'\big)\tilde{\nabla}^{a}\Theta \Big] \nonumber\\
&\qquad + \big(1 + \mathit{p}'\big) \Theta (C_6)^{a}{}_{b}\omega^{b} \label{DivC10}
\end{align}
As before, we shall extract the non-zero terms from equation \eqref{DivC10}, to obtain the following constraint $(C_{11})^{a}$ equation below, which is actually the divergence of the curl of $(C_{6})^{ab}$, we can see that the seventh term on the right hand side of equation \eqref{C11} is the term we are looking for,
\begin{align}
0=(C_{\text{11}})^{a} &:= - \tfrac{1}{6} (2 - 9 \phi) \Theta\dot{u}^{b} \tilde{\nabla}_{b}\omega^{a} - \tfrac{1}{2} (2 - \phi) \Theta \omega^{b} \tilde{\nabla}_{b}\dot{u}^{a}  \nonumber \\ 
&\qquad + \tfrac{1}{6} (1 - 9 \mathit{p}') \tilde{\nabla}^{b}\Theta\tilde{\nabla}_{b}\omega^{a} - \tfrac{1}{6} (5 + 3 \mathit{p}') \omega^{b} \tilde{\nabla}_{b}\tilde{\nabla}^{a}\Theta\nonumber \\
&\qquad - \tfrac{1}{2} (1 +  \mathit{p}' - \phi) \dot{u}^{b} \omega_{b} \tilde{\nabla}^{a}\Theta  - \tfrac{1}{9}(1 + 3 \phi)\Theta\epsilon^{a}{}_{bc} \dot{u}^{b} \tilde{\nabla}^{c}\Theta \nonumber\\
&\qquad + \Bigl[\tfrac{1}{6}(2 + 9 \phi) \Theta\mathcal{J} - \tfrac{1}{3} (1 + 3 \mathit{p}') \Theta\mathcal{E} - \tfrac{1}{2} (6 \phi +  \psi) \Theta\dot{u}_{b}\dot{u}^{b} \nonumber \\ 
&\qquad + \tfrac{1}{6} \big(19 - 18 \mathit{p}' - 81 \mathit{p}'^2\big) \Theta \omega^2 - \tfrac{1}{3} (5 - 9 \mathit{p}'- 9 \phi) \dot{u}^{b} \tilde{\nabla}_{b}\Theta\Bigr]\omega^{a}  \nonumber \\ 
&\qquad - \Bigl[\tfrac{1}{18} (8 - 36 \phi + 3 \psi) \Theta\dot{u}^{b} \omega_{b} + \tfrac{1}{6} (2 + 12 \mathit{p}'- 3 \phi) \omega^{b} \tilde{\nabla}_{b}\Theta\Bigr] \dot{u}^{a} \label{C11}
\end{align}
By contracting constraint $(C_{7})^{ab}$ in equation \eqref{C7ab}, which is the time propagation of $(C_{6})^{ab}$ with $\omega_{b}$ we obtain the following equation, again in equation \eqref{C7abOmega} one can see that the third term in on the right hand side is our term of interest,
\begin{align}
(C_7)_{b}{}^{a} \omega^{b} &= \Big[\tfrac{1}{9} \big(2 - 9 \phi + 3 \psi\big)\Theta\dot{u}^{b}\omega_{b} + \big(\tfrac{1}{3} + \mathit{p}' -  \phi\big) \omega^{b} \tilde{\nabla}_{b}\Theta\Big]\dot{u}^{a}  \nonumber \\ 
&\qquad  - \Bigl[ \tfrac{1}{9} (2 - 3 \phi)\Theta\mathcal{J} - \tfrac{1}{27}  (4 - 3 \psi)\Theta\dot{u}_{b} \dot{u}^{b} +  \bigl( \tfrac{7}{9} - 2 \mathit{p}' + 3 \mathit{p}'^2\bigr) \Theta\omega^2 \nonumber\\
&\qquad - \tfrac{2}{3}  \phi \dot{u}^{b}\tilde{\nabla}_{b}\Theta\Bigr]\omega^{a} + \big(\tfrac{1}{3} + \mathit{p}' -  \phi\big) \dot{u}^{b}\omega_{b} \tilde{\nabla}^{a}\Theta + \big(\tfrac{2}{3} -  \phi\big) \Theta\omega^{b} \tilde{\nabla}_{b}\dot{u}^{a}  \nonumber\\
&\qquad + \big(\tfrac{1}{3} + \mathit{p}'\big) \omega^{b} \tilde{\nabla}_{b}\tilde{\nabla}^{a}\Theta - \Bigl[\tfrac{1}{3} (1 + 3 \mathit{p}') (C_1)^{b}\dot{u}_{b} + \tfrac{1}{6}(1 + 3 \mathit{p}') \tilde{\nabla}_{b}(C_1)^{b} \Bigr]\omega^{a} . \label{C7abOmega} 
\end{align}
From the non-zero terms in equation \eqref{C7abOmega}, we extract the new constraint $(C_{12})^{a}$ below. Now we have what we need to be able to eliminate the double spatial gradients of the expansion scalar from constraint $(C_{7})^{ab}$ so as to retain terms only involving the acceleration vector $\dot{u}^{a}$, the vorticity $\omega^{a}$ and the spatial gradient of the vorticity vector fields $\tilde{\nabla}_{a}\omega^{b}$.
\begin{align}
0=(C_{12})^{a} &:= \Big[\tfrac{1}{9} \big(2 - 9 \phi + 3 \psi\big)\Theta\dot{u}^{b}\omega_{b} + \big(\tfrac{1}{3} + \mathit{p}' -  \phi\big) \omega^{b} \tilde{\nabla}_{b}\Theta\Big]\dot{u}^{a}  \nonumber \\ 
&\qquad  - \Bigl[ \tfrac{1}{9} (2 - 3 \phi)\Theta\mathcal{J} - \tfrac{1}{27}  (4 - 3 \psi)\Theta\dot{u}_{b} \dot{u}^{b} +  \bigl( \tfrac{7}{9} - 2 \mathit{p}' + 3 \mathit{p}'^2\bigr) \Theta\omega^2 \nonumber\\
&\qquad - \tfrac{2}{3}  \phi \dot{u}^{b}\tilde{\nabla}_{b}\Theta\Bigr]\omega^{a} + \big(\tfrac{1}{3} + \mathit{p}' -  \phi\big) \dot{u}^{b}\omega_{b} \tilde{\nabla}^{a}\Theta + \big(\tfrac{2}{3} -  \phi\big) \Theta\omega^{b} \tilde{\nabla}_{b}\dot{u}^{a}  \nonumber\\
&\qquad + \big(\tfrac{1}{3} + \mathit{p}'\big) \omega^{b} \tilde{\nabla}_{b}\tilde{\nabla}^{a}\Theta .\label{C12} 
\end{align}
Putting equation \eqref{C12}, \eqref{C11} and \eqref{C9} together yields the following constraint equation \eqref{C13} below, the advantage of this constraint will be evident in the following sections since it isolates the term $\omega^{b} \tilde{\nabla}_{b}\tilde{\nabla}^{a}\Theta$, this is not exactly what we were hoping for, but it is a step in the right direction.
\begin{align}
0=(C_{\text{13}})^{a} &:= \tfrac{1}{2}\omega^{a} \bigl[\tfrac{1}{3} (2 + 15 \phi) \Theta\mathcal{J}  - (1 + 3 \mathit{p}') \Theta\mathcal{E} + \tfrac{1}{9} (2 - 81 \phi - 15 \psi) \Theta\dot{u}^2 \nonumber \\ 
&\qquad + \tfrac{1}{3} \big(25 - 18 \mathit{p}' - 135 \mathit{p}'^2\big) \Theta \omega^2 - (5 - 9 \mathit{p}' - 10 \phi) \dot{u}^{b} \tilde{\nabla}_{b}\Theta\bigr] \nonumber \\ 
&\qquad - \tfrac{1}{4} \dot{u}^{a} \bigl[(2 - 9 \phi)\Theta\dot{u}^{b}\omega_{b} + (1 + 9 \mathit{p}') \omega^{b} \tilde{\nabla}_{b}\Theta\bigr] - \omega^{b} \tilde{\nabla}_{b}\tilde{\nabla}^{a}\Theta\nonumber \\ 
&\qquad - \tfrac{1}{2}\dot{u}^{b} \omega_{b} \tilde{\nabla}^{a}\Theta -  \Theta \omega^{b} \tilde{\nabla}_{b}\dot{u}^{a} - \tfrac{1}{4} (2 - 9 \phi) \Theta\dot{u}^{b}\tilde{\nabla}_{b}\omega^{a} \nonumber \\ 
&\qquad + \tfrac{1}{4} (1 - 9 \mathit{p}') \tilde{\nabla}^{b}\Theta\tilde{\nabla}_{b}\omega^{a} - \tfrac{1}{6}(1 + 3 \phi) \Theta\epsilon^{a}{}_{bc}\dot{u}^{b}  \tilde{\nabla}^{c}\Theta . \label{C13}
\end{align}
We have established that when $\{\sigma_{ab} = 0, T_{ab} = \rho u_a u_b + p h_{ab} \} $ the following constraints must be satisfied: $(C_{1})^a \rightarrow (C_{5})^a$ and $(C_{6})^{ab}$.  The time propagation of these constraints is identically zero if $(C_{7})^{ab}$ is true. The next level set of consistency relations $(C_{8})^a\rightarrow (C_{13})^a$ then follow. These shall be used to derive some specific case in the next section (\ref{Sec7}). One still have to compute the time propagation of equation \eqref{C13} constraint $(C_{13})^a$  in the general case to nail down the proof. 
\section{The case where the vorticity and the acceleration vector fields are parallel}\label{Sec7}
\textbf{Theorem 2:} \textit{If a rotating and expanding shear-free perfect fluid obeys a barotropic equation of state and $\dot{u}^{a} = \psi \omega^{a}$, then $\Theta\omega=0$.}\\
\newline 
\textit{Proof:} Let us define the acceleration vector field to be parallel to the vorticity vector as follows
\begin{equation}
\dot{u}^{a} = \psi \omega^{a} \label{AccParVort},
\end{equation}
where the symbol $\psi$ should not be confused with the on from the previous section, here we have chosen $\psi$ so as to mimic the calculation in \cite{senovilla1998theorems}, with the condition that $\psi\neq0$ and $\psi$ is a space-time scalar field. With this in mind, equation \eqref{PropAcc} and \eqref{OmegaDot1} can be written as two propagation equations for the acceleration vector field as follows,

\begin{equation}
\ddot{u}^{\langle a\rangle} = - \tfrac{3}{2}\mathit{p}' (C_1)^{a}  - \Big( \tfrac{1}{3} +  \tfrac{\mathcal{E} \mathit{p}^{\prime\prime}}{\mathit{p}'} - \mathit{p}' - \tfrac{3 \mathit{p}'^2}{\psi^2}\Big) \Theta\dot{u}^{a}  + \frac{3 \mathit{p}'}{2 \psi}\epsilon^{a}{}_{bc}\omega^{b} \tilde{\nabla}^{c}\psi ,
\label{PropAcc1}
\end{equation}

\begin{equation}
\ddot{u}^{\langle a\rangle} = \Bigl[\tfrac{\dot{\psi}}{\psi} - \big( \tfrac{2}{3} - \mathit{p}'\big) \Theta\Bigr]\dot{u}^{a}.
\label{PropAcc2}
\end{equation}
Comparing equation \eqref{PropAcc1} and \eqref{PropAcc2}, we can extract the following information regarding the scalar $\psi$,

\begin{equation}
\epsilon^{a}{}_{bc} \omega^{b} \tilde{\nabla}^{c}\psi = 0 \label{GradPsi1}
\end{equation}

\begin{equation}
\dot{\psi} = \Big(\tfrac{1}{3} -  \tfrac{\mathcal{E} \mathit{p}^{\prime\prime}}{\mathit{p}'} + \tfrac{3 \mathit{p}'^2}{\psi^2}\Big) \Theta\psi. \label{PsiDot}
\end{equation}
Where equation \eqref{GradPsi1}  tells us that the spatial gradient of $\psi$ is also parallel to the vorticity vector field and equation \eqref{PsiDot} is the time propagation equation for $\psi$. Applying equation \eqref{AccParVort} and \eqref{GradPsi1} to constraint $(C_1)^a$ in equation \eqref{eq:RICE:1} this leads to the following simple expression for the spatial gradient of the expansion scalar $\Theta$, where we can see that it is parallel to the vorticity vector field as well,

\begin{equation}
\tilde{\nabla}^{a}\Theta = - \tfrac{3}{2} (C_1)^{a} + \tfrac{3 \mathit{p}'}{\psi}\Theta \omega^{a}.  \label{GradTheta}
\end{equation}
When applying equation \eqref{GradTheta} in the identity \eqref{SSCofS}  one obtains the simplified evolution equation for the expansion scalar as follows,
\begin{equation}
\dot{\Theta} = \tfrac{3 \mathit{p}'^2 }{\psi^2}\Theta^2 + \tfrac{3}{4 \omega^2}\epsilon_{abd} \omega^{a} \tilde{\nabla}^{d}(C_1)^{b}. \label{ThetaDot1}
\end{equation}
On taking the spatial divergence of constraint $(C_1)^a$ in equation \eqref{eq:RICE:1} and using the following relations \eqref{GradComAcc}, \eqref{GradComOmega} and \eqref{SSDCofOmega}, we obtain the following expression

\begin{equation}
0=\tilde{\nabla}_{c}(C_1)^{c} := - \tfrac{2}{3} \tilde{\nabla}^2\Theta - \big( \tfrac{2}{3} - 6 \mathit{p}'\big) \Theta \omega^2 -  \tfrac{4}{3} \dot{u}^{c} \tilde{\nabla}_{c}\Theta - 2 (C_1)^{c} \dot{u}_{c}, \label{DivC1}
\end{equation}
Upon applying equation \eqref{GradTheta}, \eqref{AccParVort} and \eqref{eq:RICE:2} in equation \eqref{DivC1} above we obtain the following expression,
\begin{equation}
0 = - \tfrac{2\mathit{p}'}{\psi}\Theta(C_2) + \tfrac{3\mathit{p}'}{\psi}(C_1)^{c} \omega_{c} - 2\Big( \tfrac{1}{3} - \tfrac{\mathcal{E} \mathit{p}^{\prime\prime}}{\mathit{p}'} +  \tfrac{3 \mathit{p}'^2}{\psi^2}\Big) \Theta \omega^2 + \tfrac{2 \mathit{p}'}{\psi^2}\Theta \omega^{c} \tilde{\nabla}_{c}\psi. \label{DivC11}
\end{equation}
From the non-zero terms in equation  \eqref{DivC11}  we can write down the following constraint equation,

\begin{equation}
0=(C_{14}) := \Theta \Bigl[ \tfrac{\mathit{p}' }{\psi^2}\omega^{c} \tilde{\nabla}_{c}\psi-\Big( \tfrac{1}{3} - \tfrac{\mathcal{E} \mathit{p}^{\prime\prime}}{\mathit{p}'} +  \tfrac{3 \mathit{p}'^2}{\psi^2}\Big) \omega^2 \Bigr]. \label{GradPsi2}
\end{equation}
From equation \eqref{GradPsi2} we have that either $\Theta=0$, and we have the proof, or that the terms in the square brackets must vanish, which gives us the following constraint equation

\begin{equation}
0=(C_{15}) := \tfrac{\mathit{p}' }{\psi^2}\omega^{c} \tilde{\nabla}_{c}\psi-\Big( \tfrac{1}{3} - \tfrac{\mathcal{E} \mathit{p}^{\prime\prime}}{\mathit{p}'} +  \tfrac{3 \mathit{p}'^2}{\psi^2}\Big) \omega^2. \label{C15}
\end{equation}
We can also find an expression for $\omega^{c} \tilde{\nabla}_{c}\psi$ by introducing equation \eqref{ThetaDot1} into the Raychaudhuri equation \eqref{eq:RI:1} and applying the condition of the case of this section (\ref{Sec6}) as follows,
\begin{align}
0 &= \Lambda -  \tfrac{3}{2} \mathcal{E} + \mu - \Big( \tfrac{1}{3} +  \tfrac{3 \mathit{p}'^2}{\psi^2}\Big) \Theta^2 + 2(1 +  \psi^2) \omega^2 + \omega^{a} \tilde{\nabla}_{a}\psi + (C_2) \psi \nonumber\\
&\qquad -  \tfrac{3}{4 \omega^2}\epsilon_{abc} \omega^{a} \tilde{\nabla}^{c}(C_1)^{b} \label{Raychaudhuri}
\end{align}
Where from the non-vanishing terms in equation \eqref{Raychaudhuri}, we can extract the following constraint which is an expression for $\omega^{c} \tilde{\nabla}_{c}\psi$ below,

\begin{equation}
0=(C_{16}) := \Lambda -  \tfrac{3}{2} \mathcal{E} + \mu - \Big( \tfrac{1}{3} +  \tfrac{3 \mathit{p}'^2}{\psi^2}\Big) \Theta^2 + 2(1 +  \psi^2) \omega^2 + \omega^{a} \tilde{\nabla}_{a}\psi.  \label{GradPsi3}
\end{equation}
Considering equations \eqref{C15} and \eqref{GradPsi3}, we can eliminate the $\omega^a\tilde{\nabla}_a\psi$ term to obtain the following constraint equation,
\begin{align}
0&=(C_{17}) :=   - \Bigl[3 \mathit{p}'^2 \big(2 + 3 \mathit{p}'\big) \psi^2 - \Big(3 \mathcal{E} \mathit{p}^{\prime\prime} -  \mathit{p}' - 6 \mathit{p}'^2\Big) \psi^4\Bigr] \omega^2 \nonumber\\
&\qquad + \mathit{p}'^2\Big(9 \mathit{p}'^2 +  \psi^2\Big)\Theta^2  -  \tfrac{3}{2} \mathit{p}'^2 \psi^2 \big(2 \Lambda - 3 \mathcal{E} + 2 \mu\big). \label{C17}
\end{align}
Where, when operating on equation \eqref{C17} with the operator $\epsilon_{abc}\omega^b\tilde{\nabla}^c$, it yields the following constraint equation,
\begin{align}
\epsilon^{a}{}_{bc} \omega^{b} \tilde{\nabla}^{c}(C_{17}) &= 2 \psi^2\bigg[\Bigl\{3 \mathit{p}'^2 \big(2 + 3 \mathit{p}'\big) - \big(3 \mathcal{E} \mathit{p}^{\prime\prime} -  \mathit{p}' - 6 \mathit{p}'^2\big) \psi^2\Bigr\} \omega \epsilon^{a}{}_{bc}\omega^{b} \tilde{\nabla}^{c}\omega \nonumber \\ 
&\qquad - \tfrac{3}{2} \mathit{p}'^2\Big(1 + \tfrac{9 \mathit{p}'^2}{ \psi^2}\Big) \Theta \epsilon^{a}{}_{bc}(C_1)^{b}\omega^{c} \bigg]. \label{Grad1C17}
\end{align}
The non-zero terms in equation \eqref{Grad1C17} form a new constraint equation below,
\begin{align}
0=(C_{18})^a &:= \Bigl[3 \mathit{p}'^2 \big(2 + 3 \mathit{p}'\big) - \big(3 \mathcal{E} \mathit{p}^{\prime\prime} -  \mathit{p}' - 6 \mathit{p}'^2\big) \psi^2\Bigr] \omega \epsilon^{a}{}_{bc}\omega^{b} \tilde{\nabla}^{c}\omega.  \label{C18}
\end{align}
Equation \eqref{C18} presents us with three options, where either $\omega=0$ and the proof is finished, or the terms in the square brackets must vanish, giving us our first non-trivial case as follows,
\begin{itemize}
\item[Case (i):] \label{Case1}

\begin{equation}
0=(C_{19}) := 3 \mathit{p}'^2 \big(2 + 3 \mathit{p}'\big) - \big(3 \mathcal{E} \mathit{p}^{\prime\prime} - \mathit{p}' - 6 \mathit{p}'^2\big) \psi^2 , \label{C19}
\end{equation}
as a consequence of \eqref{C19}, equation \eqref{C17} gives us the following constraint as well
\begin{equation}
0=(C_{20}) := \Lambda -  \tfrac{3}{2} \mathcal{E} + \mu - \Big( \tfrac{1}{3} + \tfrac{3 \mathit{p}'^2}{\psi^2}\Big) \Theta^2 .\label{Case1:C20}
\end{equation}
Time-propagating this equation \eqref{Case1:C20} results in the following,
 
\begin{equation}
(\dot{C}_{20}) =  \tfrac{1}{2}\big(1 + 3 \mathit{p}'\big) \Theta\mathcal{E} - \tfrac{1}{2 \omega^2} \Big( 1 + \tfrac{9\mathit{p}'^2}{ \psi^2}\Big) \Theta\epsilon_{abc} \omega^{a} \tilde{\nabla}^{c}(C_1)^{b}. \label{C20Dot}
\end{equation}
Furthermore, the non-zero terms on the right hand side of equation \eqref{C20Dot}, become the following constraint equation,
\begin{equation}
0=(C_{21}) := \big(1 + 3 \mathit{p}'\big) \Theta\mathcal{E}  \label{C21}. 
\end{equation}
Where either $\Theta=0$ and the proof is done, or $p^{\prime}=-\tfrac{1}{3}$ which by equation \eqref{C17} implies that $\psi^2+1=0$ which is not allowed. Therefore $\Theta\omega=0$.

\item[Case (ii):]  or, \label{Case2}
 \begin{equation}
0=(C_{\text{22}})^{a} := \epsilon^{a}{}_{bc} \omega^{b} \tilde{\nabla}^{c}\omega. \label{Case2:C22}
\end{equation}
The time propagation of equation \eqref{C17} gives us the following expression,
\begin{align}
(\dot{C}_{17}) &= \Theta \biggl\{-\mathcal{E} \Bigl[9 \mathit{p}'^3 (4 + 3 \mathit{p}') +  \tfrac{9}{2} \mathit{p}'^2 (1 + 9 \mathit{p}') \psi^2\Bigr] + \big(\Lambda +\mu\big) \Bigl[6 \mathit{p}'^3 (4 + 3 \mathit{p}') \nonumber \\ 
&\qquad + 2 \mathit{p}'^2 (1 + 12 \mathit{p}') \psi^2\Bigr] + \Bigl[-4 \mathit{p}'^3 (2 + 3 \mathit{p}' + 18 \mathit{p}'^2) -  \tfrac{18 \mathit{p}'^5}{\psi^2}(4 + 3 \mathit{p}') \nonumber \\ 
&\qquad -  \tfrac{2}{3} \mathit{p}'^2 (1 + 12 \mathit{p}') \psi^2\Bigr] \Theta^2 + \Bigl[27 \mathit{p}'^4 (2 + 3 \mathit{p}') + 3 \mathit{p}'^3 (7 + 15 \mathit{p}') \psi^2 \nonumber \\ 
&\qquad  + \tfrac{1}{3} \mathit{p}' (1 + 6 \mathit{p}') (-2 + 9 \mathit{p}') \psi^4\Bigr] \omega^2 -  \tfrac{4 }{3 \mathit{p}'}\mathcal{Z}^2 \omega^2 - 3 \mathcal{E}^2 \mathit{p}^{(3)} \psi^4 \omega^2\nonumber \\ 
&\qquad + \mathcal{Z} \Bigl[4\mathit{p}' \big(\Lambda  - \tfrac{3}{2} \mathcal{E} \mathit{p}' +  \mu\big) - 4\mathit{p}' \Big( \tfrac{1}{3}  +  \tfrac{3 \mathit{p}'^2}{\psi^2}\Big) \Theta^2 - \Bigl(\mathit{p}' (8 - 3 \mathit{p}')\nonumber \\ 
&\qquad + (2 + 3 \mathit{p}') \psi^2\Bigr) \omega^2\Bigr] + \tfrac{3\mathit{p}'^2}{2\omega^2} \big(9 \mathit{p}'^2 + \psi^2\big) \epsilon_{abc}\omega^{a} \tilde{\nabla}^{c}(C_1)^{b}\biggr\}. \label{Case2:C17Dot}
\end{align}
Where for simplicity we have defined $\mathcal{Z}$ to be the following combination since it appears repeatedly in the following calculations,
\begin{equation}
\mathcal{Z} = 3 \mathit{p}'^2 \big(2 + 3 \mathit{p}'\big) - \big(3 \mathcal{E} \mathit{p}^{\prime\prime} - \mathit{p}' - 6 \mathit{p}'^2\big) \psi^2 . \label{Case1:C19}
\end{equation}
From the non-zero terms in equation \eqref{Case2:C17Dot}, we can deduce that either $\Theta=0$ and the proof is done, or the non-zero terms in the curly brackets vanish, which gives us the following constraint,
\begin{align}
0=(C_{23}) &:= -\mathcal{E} \Bigl[9 \mathit{p}'^3 (4 + 3 \mathit{p}') +  \tfrac{9}{2} \mathit{p}'^2 (1 + 9 \mathit{p}') \psi^2\Bigr] + \big(\Lambda +\mu\big) \Bigl[6 \mathit{p}'^3 (4 + 3 \mathit{p}') \nonumber \\ 
&\qquad + 2 \mathit{p}'^2 (1 + 12 \mathit{p}') \psi^2\Bigr] + \Bigl[-4 \mathit{p}'^3 (2 + 3 \mathit{p}' + 18 \mathit{p}'^2) -  \tfrac{18 \mathit{p}'^5}{\psi^2}(4 + 3 \mathit{p}') \nonumber \\ 
&\qquad -  \tfrac{2}{3} \mathit{p}'^2 (1 + 12 \mathit{p}') \psi^2\Bigr] \Theta^2 + \Bigl[27 \mathit{p}'^4 (2 + 3 \mathit{p}') + 3 \mathit{p}'^3 (7 + 15 \mathit{p}') \psi^2 \nonumber \\ 
&\qquad  + \tfrac{1}{3} \mathit{p}' (1 + 6 \mathit{p}') (-2 + 9 \mathit{p}') \psi^4\Bigr] \omega^2 -  \tfrac{4 }{3 \mathit{p}'}\mathcal{Z}^2 \omega^2 - 3 \mathcal{E}^2 \mathit{p}^{(3)} \psi^4 \omega^2\nonumber \\ 
&\qquad + \mathcal{Z} \Bigl[4\mathit{p}' \big(\Lambda  - \tfrac{3}{2} \mathcal{E} \mathit{p}' +  \mu\big) - 4\mathit{p}' \Big( \tfrac{1}{3}  +  \tfrac{3 \mathit{p}'^2}{\psi^2}\Big) \Theta^2 - \Bigl(\mathit{p}' (8 - 3 \mathit{p}')\nonumber \\ 
&\qquad + (2 + 3 \mathit{p}') \psi^2\Bigr) \omega^2\Bigr] . \label{Case2:C23}
\end{align}
Furthermore, when applying the operator $\omega^a\tilde{\nabla}_a$ in equation \eqref{C17} one obtains the following expression,
\begin{align}
\omega^{a} \tilde{\nabla}_{a}(C_{17})&= \tfrac{\psi}{\mathit{p}'}\omega^2\biggl\{-\mathcal{E} \Bigl[9 \mathit{p}'^3 (4 + 3 \mathit{p}') +  \tfrac{9}{2} \mathit{p}'^2 (1 + 9 \mathit{p}') \psi^2\Bigr] + \big(\Lambda +\mu\big) \Bigl[6 \mathit{p}'^3 (4 + 3 \mathit{p}') \nonumber \\ 
&\qquad + 2 \mathit{p}'^2 (1 + 12 \mathit{p}') \psi^2\Bigr] + \Bigl[-4 \mathit{p}'^3 (2 + 3 \mathit{p}' + 18 \mathit{p}'^2) -  \tfrac{18 \mathit{p}'^5}{\psi^2}(4 + 3 \mathit{p}') \nonumber \\ 
&\qquad -  \tfrac{2}{3} \mathit{p}'^2 (1 + 12 \mathit{p}') \psi^2\Bigr] \Theta^2 + \Bigl[27 \mathit{p}'^4 (2 + 3 \mathit{p}') + 3 \mathit{p}'^3 (7 + 15 \mathit{p}') \psi^2 \nonumber \\ 
&\qquad  + \tfrac{1}{3} \mathit{p}' (1 + 6 \mathit{p}') (-2 + 9 \mathit{p}') \psi^4\Bigr] \omega^2 -  \tfrac{4 }{3 \mathit{p}'}\mathcal{Z}^2 \omega^2 - 3 \mathcal{E}^2 \mathit{p}^{(3)} \psi^4 \omega^2\nonumber \\ 
&\qquad + \mathcal{Z} \Bigl[4\mathit{p}' \big(\Lambda  - \tfrac{3}{2} \mathcal{E} \mathit{p}' +  \mu\big) - 4\mathit{p}' \Big( \tfrac{1}{3}  +  \tfrac{3 \mathit{p}'^2}{\psi^2}\Big) \Theta^2 - \Bigl(\mathit{p}' (8 - 3 \mathit{p}')\nonumber \\ 
&\qquad + \tfrac{1}{3}(2 + 15 \mathit{p}') \psi^2\Bigr) \omega^2 + \tfrac{2 \mathit{p}' \psi}{\omega}\omega^{a} \tilde{\nabla}_ {a}\omega\Bigr] - \tfrac{3\mathit{p}'^3}{\omega^2}\Big(\tfrac{9 \mathit{p}'^2}{\psi} + \psi\Big) \Theta (C_ 1)^{a}\omega_{a} \nonumber\\
&\qquad + (C_{15})\psi^2 \Bigl[6 \mathit{p}'^2 \big(2 + 3 \mathit{p}'\big) + 4 \mathcal{Z} - \tfrac{\mathit{p}'^2}{\omega^2}\big(6 \Lambda -9 \mathcal{E} - 2 \Theta^2 + 6\mu\big)\Bigr]\bigg\} . \label{Case2:Grad2C17}
\end{align}
From equation \eqref{Case2:Grad2C17} above we can see that either $\omega=0$ and we have the proof, or the non-zero terms within the curly brackets  must vanish , giving us the following constraint equation,
\begin{align}
0=(C_{24})&:= -\mathcal{E} \Bigl[9 \mathit{p}'^3 (4 + 3 \mathit{p}') +  \tfrac{9}{2} \mathit{p}'^2 (1 + 9 \mathit{p}') \psi^2\Bigr] + \big(\Lambda +\mu\big) \Bigl[6 \mathit{p}'^3 (4 + 3 \mathit{p}') \nonumber \\ 
&\qquad + 2 \mathit{p}'^2 (1 + 12 \mathit{p}') \psi^2\Bigr] + \Bigl[-4 \mathit{p}'^3 (2 + 3 \mathit{p}' + 18 \mathit{p}'^2) -  \tfrac{18 \mathit{p}'^5}{\psi^2}(4 + 3 \mathit{p}') \nonumber \\ 
&\qquad -  \tfrac{2}{3} \mathit{p}'^2 (1 + 12 \mathit{p}') \psi^2\Bigr] \Theta^2 + \Bigl[27 \mathit{p}'^4 (2 + 3 \mathit{p}') + 3 \mathit{p}'^3 (7 + 15 \mathit{p}') \psi^2 \nonumber \\ 
&\qquad  + \tfrac{1}{3} \mathit{p}' (1 + 6 \mathit{p}') (-2 + 9 \mathit{p}') \psi^4\Bigr] \omega^2 -  \tfrac{4 }{3 \mathit{p}'}\mathcal{Z}^2 \omega^2 - 3 \mathcal{E}^2 \mathit{p}^{(3)} \psi^4 \omega^2\nonumber \\ 
&\qquad + \mathcal{Z} \Bigl[4\mathit{p}' \big(\Lambda  - \tfrac{3}{2} \mathcal{E} \mathit{p}' +  \mu\big) - 4\mathit{p}' \Big( \tfrac{1}{3}  +  \tfrac{3 \mathit{p}'^2}{\psi^2}\Big) \Theta^2 - \Bigl(\mathit{p}' (8 - 3 \mathit{p}')\nonumber \\ 
&\qquad + \tfrac{1}{3}(2 + 15 \mathit{p}') \psi^2\Bigr) \omega^2 + \tfrac{2 \mathit{p}' \psi}{\omega}\omega^{a} \tilde{\nabla}_ {a}\omega\Bigr].  \label{Case2:C24}
\end{align} 

Now, comparing equation \eqref{Case2:C23} and \eqref{Case2:C24}, we arrive at the following simple looking constraint equation shown below,
\begin{equation}
0 = (C_{25}) := \mathcal{Z} \Bigl[\big(\tfrac{2}{3} -  \mathit{p}'\big) \omega^3 + \tfrac{\mathit{p}'}{\psi}\omega^{a} \tilde{\nabla}_{a}\omega\Bigr]. \label{Case2:C25}
\end{equation}
Since $\mathcal{Z}$ cannot be zero in this case, only the terms within the square brackets in equation \eqref{Case2:C25} must vanish, which leads us to the following constraint,
\begin{equation}
0 = (C_{26}) :=\big(\tfrac{2}{3} -  \mathit{p}'\big) \omega^3 + \tfrac{\mathit{p}'}{\psi}\omega^{a} \tilde{\nabla}_{a}\omega. \label{Case2:C26}
\end{equation}
Equation \eqref{Case2:C22} and \eqref{Case2:C26}, become new constraints which require our attention. The time propagation of \eqref{Case2:C22} gives us the following,
\begin{align}
\dot{\left( C_{\text{22}}\right)}^{\langle a\rangle} &= \big( 2 \mathit{p}'- \tfrac{5}{3} \big) \Theta(C_{\text{22}})^{a} + \tfrac{\psi }{\mathit{p}'}(C_{26})\omega^{a} +   \big(\tfrac{3}{2} \mathit{p}'-1\big) \omega \epsilon^{a}{}_{bc}(C_1)^{b}\omega^{c} \nonumber\\
&\qquad + \big(1 -  \tfrac{2}{3 \mathit{p}'}\big) \psi \omega^3 \omega^{a} -  \omega^2 \tilde{\nabla}^{a}\omega .\label{Case2:C22Dot}
\end{align} 
The non-zero terms in equation \eqref{Case2:C22Dot}  gives us the following new constraint equation below,
\begin{equation}
0=(C_{\text{27}})^{a} := \omega^2 \Bigl[\big(1 -  \tfrac{2}{3 \mathit{p}'}\big) \psi \omega \omega^{a} -  \tilde{\nabla}^{a}\omega\Bigr]. \label{Case2:C27}
\end{equation}
Equation \eqref{Case2:C27} implies that either the vorticity scalar $\omega=0$ where the proof is done, or that the terms within the square brackets must vanish, giving us the following constraint,
\begin{equation}
0=(C_{\text{28}})^{a} := \big(1 -  \tfrac{2}{3 \mathit{p}'}\big) \psi \omega \omega^{a} -  \tilde{\nabla}^{a}\omega .\label{Case2:C28}
\end{equation}
Equation \eqref{Case2:C28} is close under time propagation and does not give us any new information, see equation \eqref{Case2:C28Dot} below,
\begin{equation}
\dot{(C_{\text{28}})}^{\langle a\rangle} = \big(\mathit{p}'- 1\big) \Theta(C_{\text{28}})^{a} -   \epsilon^{a}{}_{bc}(C_{\text{28}})^{b} \omega^{c} + \big(\tfrac{3}{2} \mathit{p}'- 1\big) \omega(C_1)^{a} . \label{Case2:C28Dot}
\end{equation}
On the other hand, equation \eqref{Case2:C26} closes under time-propagation and does not give us any new information, see equation\eqref{Case2:C26Dot} below,
\begin{equation}
(\dot{C}_{26}) = \Big(-2 + 2 \mathit{p}' -  \tfrac{3 \mathit{p}'^2}{\psi^2}\Big) \Theta (C_{26})+  \tfrac{\mathit{p}'}{\psi}\Big(1 -  \tfrac{3 \mathit{p}'}{2}\Big) \omega (C_1)^{a}\omega_{a} . \label{Case2:C26Dot}
\end{equation}
At this point we shall consider the Ricci identities for the vorticity given by equation\eqref{SSDCofOmega},  together with all the relations that we have obtained up until now, we arrive at the following constraint,
\begin{align}
0 &= (C_{\text{28}})_{a} (C_{\text{28}})^{a} - \tfrac{2}{\mathit{p}'}\big( 1 + \tfrac{1}{3 \mathit{p}'}\big) \psi^2 \omega(C_{26}) +  \tfrac{2}{3}\big(2 -  \tfrac{1}{ \mathit{p}'}\big) \psi \omega^2(C_2) \nonumber \\ 
&\qquad - \frac{2}{3 \mathit{p}'}\big(1 + \tfrac{1}{ \mathit{p}'}\big) \psi^2 \omega^2(C_{15}) - \big( \tfrac{2}{3} (C_{18}) +  \mathcal{E} - \tfrac{2 \mathit{p}'^2 }{\psi^2}\Theta^2\big) \omega^2 + \Bigl(- \tfrac{2}{9 \mathit{p}'^2}(C_{17}) \nonumber \\ 
&\qquad + \big(1 + \tfrac{2}{3 \mathit{p}'^2} -  \tfrac{4}{3 \mathit{p}'}\big) \psi^2\Bigr) \omega^4 -2 \big(1 - \tfrac{2}{3 \mathit{p}'}\big) \psi \omega (C_{\text{28}})^{a}\omega_{a} + (C_6)_{ab} \omega^{a} \omega^{b} \nonumber \\ 
&\qquad -  \omega^{a} \tilde{\nabla}_{a}(C_2) -  \omega \tilde{\nabla}_{a}(C_{\text{28}})^{a} -  \tilde{\nabla}_{b}\omega_{a} \tilde{\nabla}^{b}\omega^{a} . \label{Case2:RicciIdent}
\end{align}
From equation \eqref{Case2:RicciIdent} one can extract the non-zero terms which form a new constraint equation below,
\begin{equation}
0 = (C_{29}) := \Big(- \mathcal{E} + \tfrac{2 \mathit{p}'^2}{\psi^2}\Theta^2\Big) \omega^2 + \Big(1 + \tfrac{2}{3 \mathit{p}'^2} -  \tfrac{4}{3 \mathit{p}'}\Big) \psi^2 \omega^4 -  \tilde{\nabla}_{b}\omega_{a} \tilde{\nabla}^{b}\omega^{a}.  \label{Case2:C29}
\end{equation}
Equation \eqref{Case2:C29} is equivalent to equation (79) in \cite{senovilla1998theorems}, where we have defined  the projection tensor $Q_{ab}$, which is orthogonal to both $u^a$ and $\omega^a$ as follows,
\begin{equation}
Q_{ab}=h_{ab}-\hat{\omega}_a\hat{\omega}_b,  \label{ProjOmega}
\end{equation}
where the $\hat{\omega}_a$ is a unit vector in the direction of the vorticity, with the property that $\dot{\hat{\omega}}^{\langle a\rangle}=0$.\\
\newline 
Finally, the time propagation of equation \eqref{Case2:C29} yields the following equation below,
\begin{align}
(\dot{C}_{29}) &= \mathcal{E} \big(\tfrac{1}{3} + \mathit{p}'\big) \Theta \omega^2 - 2\big(1 - \mathit{p}'\big) \Theta(C_{29}) + \big(2 - 3 \mathit{p}'\big) \omega(C_1)^{a} (C_{\text{28}})_{a}   \nonumber \\ 
&\qquad - 2 \Big(\tfrac{\mathit{p}'}{\psi} +  \tfrac{1}{3} \psi\Big) \Theta \omega^2(C_2) - 2 \Big(\tfrac{5 \mathit{p}'}{\psi} +   \psi\big(\tfrac{4}{3} + \mathit{p}'\big) +  \tfrac{\mathcal{Z}}{3 \mathit{p}' \psi}\Big) \Theta \omega (C_{\text{28}})^{a}\omega_{a} \nonumber \\ 
&\qquad + \Bigl((C_2) - \big(3 - \tfrac{4}{3 \mathit{p}'} - 3 \mathit{p}'\big) \psi \omega^2\Bigr) (C_1)^{a}\omega_{a} -  (C_1)^{a} \omega^{b} \tilde{\nabla}_{b}\omega_{a}\nonumber \\ 
&\qquad -  \tfrac{2}{3} (C_2) \epsilon_{abc} \omega^{a} \tilde{\nabla}^{c}\omega^{b} + \tfrac{3\mathit{p}'^2 }{\psi^2}\Theta \epsilon_{abc}\omega^{a} \tilde{\nabla}^{c}(C_1)^{b}. \label{Case2:C29Dot}
\end{align}

From equation \eqref{Case2:C29Dot} above, we can extract a new constraint from the non-zero terms as follows,
\begin{equation}
0 = (C_{30}) := \mathcal{E} \big(\tfrac{1}{3} + \mathit{p}'\big) \Theta \omega^2. \label{Case2:C30}
\end{equation}

As we have seen in case (i), this constraint leads to the required condition to prove the theorem, hence we have that $\Theta \omega^2=0$. This finishes the proof. The discussion of this section follows that of section (4) in \cite{senovilla1998theorems}, but with more details included.
\end{itemize}

\section{The case where the acceleration vector field is orthogonal to the vorticity}\label{Sec8}
In this section we present a covariant proof of the theorem that was given in \cite{van2016shear}, which was demonstrated using the orthonormal tetrad formalism.\\
\newline
\textbf{Definition:} \textit{A spatially projected tensorial object $\zeta$ on a manifold $(\mathcal{M},\bold{g},\bold{u})$ is called $\textbf{basic}$ if its spatially projected Lie derivative along $\bf{u}$ is zero \cite{slobodeanu2014shear}.}\\
\newline
\textbf{Theorem 3:} \textit{If a rotating and expanding shear-free perfect fluid obeys a barotropic equation of state and $\dot{U}^{a} \hat{\omega}_{a}$ is basic, then there exist a Killing vector parallel to the vorticity.}\\
\newline
\textit{Proof:} In the following discussion will shall assume that $\Theta\omega$ and $p^{\prime}$ are non-zero, then we will show that this leads to an inconsistency unless there exist a Killing vector along the vorticity. The condition that $\dot{U}^{a} \hat{\omega}_{a}$ is basic, where we have defined the rescaled acceleration vector $\dot{U}^{a}$ as follows, $\dot{U}^{a} = \frac{\dot{u}^{a}}{\lambda \mathit{p}'}$ with $\log\lambda=\int\tfrac{d\mu}{3\mathcal{E}}$, can be expressed in the following way,
\begin{equation}
\mathcal{L}_{u}\big[\dot{U}^{b} \hat{\omega}_{b}\big] =u^{a}\nabla_a\big[\dot{U}^{b} \hat{\omega}_{b}\big] = 0 . \label{BasicRule}
\end{equation}
And of course,  $\hat{\omega}^{a} = \frac{\omega^{a}}{\omega}$ is a unit vorticity vector. Expanding the left hand side of equation \eqref{BasicRule}, and applying the evolution equation for the acceleration \eqref{UaDot} and using the fact that $u^{c}\nabla_{c}\hat{\omega}^{a}=0$, one obtains the following equation,
\begin{equation}
\mathit{p}' \dot{U}^{a} \hat{\omega}_{a} \Theta + \hat{\omega}^{a} Z_{a} \omega = 0 .  \label{Fawa}
\end{equation}
 Where we have defined $Z^{a} = \frac{\tilde{\nabla}^{a}\Theta}{\lambda \omega}$, for reasons that will be clear later (where $Z^{a}$ has been defined such that we can construct a basic variable $ \hat{\omega}^{a} Z_{a}$).
Time propagating equation \eqref{Fawa}, results in the following constraint equation below, 
\begin{equation}
\dot{U}^{a} \hat{\omega}_{a} \mathit{p}'\left[\Lambda -  \tfrac{3}{2} \mathcal{E} + \mathcal{J} + \big(\tfrac{2}{3}  -  \phi - 2 \mathit{p}'\big) \Theta^2 +  \mu + 2 \omega^2\right] = 0 . \label{Constr01}
\end{equation}
Where it follows that either $\dot{U}^{a} \hat{\omega}_{a}=0$, or the terms in the square brackets must vanish.
In what follows, we shall establish that $\dot{U}^{a} \hat{\omega}_{a}\neq0$ is inconsistent with $\Theta\omega\neq0$. Thus, when the terms in the square parenthesis in \eqref{Constr01} vanish, we are lead to the following energy equation,  
\begin{equation}
 \Lambda -  \tfrac{3}{2} \mathcal{E} + \mathcal{J} + \left(\tfrac{2}{3} -  \phi - 2 \mathit{p}'\right) \Theta^2 + \mu + 2 \omega^2 = 0 \label{EnergyEquation}
\end{equation}
which is equivalent to equation (48) in \cite{van2016shear}.
Taking the spatial directional covariant derivative along the vorticity (implies the operator $\hat{\omega}_{a}\tilde{\nabla}^{a}$) of equation \eqref{EnergyEquation} we obtain the following constraint equation,
\begin{equation}
 \mathit{p}' \dot{U}^{a} \hat{\omega}_{a} \left(1 - 3 \phi - 3 \mathit{p}' +  \mathcal{X}\right) \Theta^2 = 0, \label{Constr02}
\end{equation}
with $\mathcal{X}$ is given by equation \eqref{XScalar}. Now, from equation \eqref{Constr02}, it is clear that the terms in the brackets must be zero, and this leads to the following equation,
\begin{equation}
1 - 3 \phi - 3 \mathit{p}' +  \mathcal{X} = 0, \label{EoSDE01}
\end{equation}
which is a 3rd order differential equation for $p$, this is equivalent to equation (51) in reference\cite{van2016shear}. Henceforth, we shall define  two \textit{force}(this is just a name tag for $\Theta\tilde{\nabla}^{a}\xi$) vector fields $F^{a}$ and $\overset{*}{F}^{a}$, which are the sum and the difference of the spatial gradient of the expansion  and the  acceleration vector multiplied by the expansion scalar as follows,
\begin{equation}
F^{a} = \lambda \left(Z^{a} \omega + \mathit{p}' \dot{U}^{a} \Theta \right), \label{ForceVector01}
\end{equation}
\begin{equation}
\overset{*}{F}^{a}= \lambda \left(Z^{a} \omega-\mathit{p}' \dot{U}^{a} \Theta \right). \label{ForceVector012}
\end{equation}
This \textit{force} vector can be written in terms of a spatial gradient of some scalar field as follows, $F^{a}=\Theta\tilde{\nabla}^{a}\xi$, where $\xi$ has been defined to be $\xi=\log \Theta-\gamma$, with $\gamma=\int\tfrac{dp}{\mathcal{E}}$.
From equation \eqref{Fawa} it follows that $F^{a}$ is orthogonal to the vorticity, and whenever $F^{a}$ vanishes, then $\nabla^{a}\xi=-u^{a}\dot{\xi}$, which shows that $u^{a}$ is hypersurface orthogonal (and hence $\omega^{a}$ must be zero) , unless $\xi$ is constant, then it follows that  $\Theta=\Theta(\mu)$ ,this case has been dealt with in \cite{treciokas1971isotropic,Lang:1993:CSG:920627,maartens1998gravito} and hence contradicts our initial assumption. A similar argument applies for $\overset{*}{F}^{a}$ in equation \eqref{ForceVector012}. Taking the spatial gradient of equation \eqref{EnergyEquation}, and using equation \eqref{EoSDE01} to eliminate $\mathcal{X}$, and applying the \textit{force} vector given in equation \eqref{ForceVector01} , we arrive at the following equation for the \textit{force} vector given in terms of the rotated acceleration vector,
\begin{equation}
F^{a} \left(\tfrac{4}{3} - 2 \phi - 3 \mathit{p}'\right) \Theta - \omega^{a}{}_{b}F^{b} \left( \tfrac{1}{2} - \tfrac{9}{2} \mathit{p}'\right) = \tfrac{9}{2}\omega^{a}{}_{b}\dot{U}^{b} \mathcal{E} \lambda \mathit{p}^{\prime\prime} \Theta . \label{ForceVector02}
\end{equation}
Where the tensor $\omega_{ab}$ has been defined as $\epsilon_{abc}\omega^{c}$ is the rotation tensor. Time propagating equation \eqref{ForceVector02}, and making use of the equation itself to eliminate terms involving $\omega^{a}{}_{b}\dot{U}^{b}$, one arrives at the following \textit{force} vector constraint equation,
\begin{equation}
F^{a} \left( \tfrac{10}{9} - \tfrac{11}{3} \phi + 3 \phi^2 - 5 \mathit{p}' + 9 \phi \mathit{p}' + 6 \mathit{p}'^2\right) \Theta  =  \omega^{a}{}_{c}F^{c} \left(\tfrac{1}{3} -  \tfrac{1}{2} \phi - 2 \mathit{p}'\right). \label{ForceVector03}
\end{equation}
Furthermore, contracting equation \eqref{ForceVector03} with the vector $\omega_{ac}F^{c} $ to annihilate the term on the left hand side, and applying equation \eqref{ProjQab} we obtain the following constraint equation
\begin{equation}
F_{a} F^{a} \left(\tfrac{2}{3} -  \phi - 4 \mathit{p}'\right) \omega^2 =  \mathcal{F}^2 \lambda^2 \left( \tfrac{2}{3} - \phi - 4 \mathit{p}'\right) \omega^2, \label{Constr03a}
\end{equation}
whereby using the fact that equation \eqref{Fawa} implies that $\hat{\omega}_{a}F^{a}=\lambda\mathcal{F}=0$, we obtain the following scalar constraint equation as follows, 
\begin{equation}
F_{a} F^{a} \left(\tfrac{2}{3} -  \phi - 4 \mathit{p}'\right) \omega^2 = 0. \label{Constr03}
\end{equation}
Equation \eqref{Constr03} clearly shows that either $F_{a} F^{a}=0$, or the terms in the brackets must vanish, where the former is inconsistent with $\Theta\omega\neq0$, and the latter implies a linear equation of state, since the time propagation of $\tfrac{2}{3} -  \phi - 4 \mathit{p}' = 0$ leads to $\mathit{p}^{\prime\prime} = 0$, where by the same argument again equation \eqref{ForceVector02} leads to $F_{a} F^{a}=0$.\\
\newline
Now let us turn to the case where $\dot{U}^{a} \hat{\omega}_{a}=0$, which by equation \eqref{Fawa} implies that $Z^{a} \hat{\omega}_{a}$ and $F^{a} \hat{\omega}_{a}$ are both zero as well.  In this situation, it is convenient to rewrite our main constraint equations \eqref{C9}, \eqref{C10}, \eqref{C12} and \eqref{C13} using the variables defined in \eqref{UScalar} to \eqref{XScalar} for transparency since most of them are zero in this case,
as follows,
\begin{align}
(C_{\text{9}})^{a} & = H^{a}{}_{b}F^{b}  \left(\tfrac{1}{9} -  \mathit{p}'\right) + H^{a}{}_{b}\dot{U}^{b} \mathcal{E} \lambda \mathit{p}^{\prime\prime} \Theta - \tfrac{1}{9}\epsilon^{a}{}_{bc} F^{b}\dot{u}^{c} \left( \tfrac{1}{3} +  \phi\right) \Theta \nonumber \\ 
&\qquad  +  \omega^{a}\Bigl[F^{b}\dot{u}_{b} \left(\tfrac{2}{3} -  \tfrac{13}{9} \phi - 2 \mathit{p}'\right)  + \Bigl\{\tfrac{1}{9}\mathcal{J} \left(\tfrac{1}{3} -  \tfrac{13}{2} \phi\right) + \tfrac{1}{3}\mathcal{E} \left(\tfrac{1}{3} +  \mathit{p}'\right)\nonumber \\ 
&\qquad + \dot{u}^2 \left(2\tfrac{ \mathcal{E} \mathit{p}^{\prime\prime}}{\mathit{p}'} -  \tfrac{13}{18} \mathcal{X}\right) - \left( \tfrac{29}{54} +  \tfrac{1}{3} \mathit{p}' - \tfrac{13}{2} \mathit{p}'^2\right) \omega^2\Bigr\}\Theta\Bigr] -  \tilde{\nabla}^{a}\mathcal{H}\nonumber \\ 
&\qquad  - \tfrac{1}{2}F^{a} \lambda \mathit{p}'\left( \tfrac{1}{9}  -  \mathit{p}'\right) \mathcal{U} \omega + \dot{u}^{a} \Bigl[ \tfrac{1}{2}\mathcal{F} \lambda \left(\tfrac{1}{9} -   \mathit{p}'\right)\omega -2 \mathcal{H}\Big] ,  \label{NewConstr9}
\end{align}
\begin{align}
(C_{\text{10}})^{a} &=  H^{a}{}_{b}F^{b} \left(1 -  \mathit{p}'\right) + H^{a}{}_{b}\dot{U}^{b}  \mathcal{E} \lambda \mathit{p}^{\prime\prime} \Theta + \epsilon^{a}{}_{bc} F^{b}\dot{u}^{c} \left(\tfrac{1}{3} -  \phi\right) \Theta \nonumber \\ 
&\qquad + \omega^{a}\Bigl[F^{b}\dot{u}_{b} \left(2 - 3 \phi - 6 \mathit{p}'\right) - \big\{\mathcal{J} \left( \tfrac{1}{3} - \tfrac{3}{2} \phi\right) + \mathcal{E} \left(\tfrac{1}{3} + \mathit{p}'\right)\nonumber \\ 
&\qquad - 3\dot{u}^2 \bigl(2\tfrac{ \mathcal{E} \mathit{p}^{\prime\prime}}{\mathit{p}'} +  \tfrac{1}{2} \mathcal{X}\bigr) - \bigl(\tfrac{19}{6} - 3 \mathit{p}' - \tfrac{27}{2} \mathit{p}'^2\bigr)\omega^2\big\}\Theta \Bigr] - \tilde{\nabla}^{a}\mathcal{I} \nonumber \\ 
&\qquad - \tfrac{3}{2}F^{a} \lambda \mathit{p}'\left( 1 - \mathit{p}'\right) \mathcal{U} \omega - \dot{u}^{a} \Bigl[\tfrac{1}{2}\mathcal{F} \lambda \left(13 +  3 \mathit{p}'\right) \omega + 10 \mathcal{I}\Bigr], \label{NewConstr10}
\end{align}
\begin{align}
(C_{\text{12}})^{a} &= H^{a}{}_{b}F^{b} \left(\tfrac{1}{3} + \mathit{p}'\right) -  H^{a}{}_{b}\dot{U}^{b} \mathcal{E} \lambda \mathit{p}^{\prime\prime} \Theta + \tfrac{1}{3}\epsilon^{a}{}_{bc} F^{b}\dot{u}^{c} \left(\tfrac{2}{3} - \phi\right) \Theta \nonumber \\ 
&\qquad + \tfrac{1}{3}\omega^{a}\Bigl[2 F^{b} \phi \dot{u}_{b} + \big\{\dot{u}^2 \mathcal{X} - \mathcal{J} \left(\tfrac{2}{3} -  \phi\right) - \left( \tfrac{7}{3} - 6 \mathit{p}' + 9 \mathit{p}'^2\right) \omega^2\big\}\Theta\Bigr]\nonumber\\
&\qquad -  \tfrac{2}{3} F^{a} \lambda \mathit{p}' \mathcal{U} \omega - \tilde{\nabla}^{a}\mathcal{G} - 2 \dot{u}^{a} \left(\tfrac{1}{3} \mathcal{F} \lambda \omega + 2 \mathcal{G}\right) \label{NewConstr12}
\end{align}
and
\begin{align}
(C_{\text{13}})^{a} &= H^{a}{}_{b}F^{b} + \epsilon^{a}{}_{bc} F^{b} \dot{u}^{c}\left(\tfrac{5}{12} -  \phi\right)  \Theta + \tfrac{1}{2}\omega^{a}\Bigl[F^{b}\dot{u}_{b} \big(3 -  \tfrac{7}{2} \phi -  9 \mathit{p}'\big)\nonumber \\ 
&\qquad - \big\{\tfrac{1}{2}\mathcal{J} \left(\tfrac{5}{3} +  \tfrac{7}{2} \phi\right)  - \tfrac{1}{2}\mathcal{E} \left(1 + 3 \mathit{p}'\right) - \dot{u}^2 \bigl(9\tfrac{\mathcal{E} \mathit{p}^{\prime\prime}}{\mathit{p}'} -  \tfrac{7}{4} \mathcal{X}\bigr) + \tfrac{1}{2} \bigl( \tfrac{71}{6}\nonumber \\ 
&\qquad - 15 \mathit{p}' - \tfrac{63}{2} \mathit{p}'^2\bigr) \omega^2\big\}\Theta\Bigr] - \tfrac{1}{8}F^{a} \lambda\mathit{p}' \left( 13 - 9 \mathit{p}'\right) \mathcal{U} \omega - \tfrac{9}{4} \bigl(\tilde{\nabla}^{a}\mathcal{G}\nonumber\\
&\qquad  + \tilde{\nabla}^{a}\mathcal{H}\bigr) - \dot{u}^{a} \Bigl[ \tfrac{1}{8}\mathcal{F} \lambda \left( 11 +  9 \mathit{p}'\right) \omega + 9 \left(\mathcal{G} +  \tfrac{1}{2} \mathcal{H}\right)\Bigr]. \nonumber \\ 
\label{NewConstr13} 
\end{align}
Therefore, since $\mathcal{U}$ and $\mathcal{F}$ are both zero, and hence $\mathcal{G}$, $\mathcal{H}$ and $\mathcal{I}$ are also zero. Equation \eqref{NewConstr13} can be written in the following way below,
\begin{align}
0= (C_{\text{13}})^{a} &:= H^{a}{}_{b}F^{b} + \epsilon^{a}{}_{bc} F^{b} \dot{u}^{c} \left(\tfrac{5}{12} -  \phi\right) \Theta + \tfrac{1}{2}\omega^{a}\Big[F^{b}\dot{u}_{b} \left(3 -  \tfrac{7}{2} \phi -  9 \mathit{p}'\right) \nonumber \\ 
&\qquad - \Theta\bigl\{\mathcal{J} \left( \tfrac{5}{6} +  \tfrac{7}{4} \phi\right) - \mathcal{E} \left(\tfrac{1}{2} + \tfrac{3}{2} \mathit{p}'\right) - \dot{u}^2 \bigl(9\tfrac{\mathcal{E}\mathit{p}^{\prime\prime}}{ \mathit{p}'} -  \tfrac{7}{4} \mathcal{X}\bigr) \nonumber \\ 
&\qquad  + \left( \tfrac{71}{12} - \tfrac{15}{2} \mathit{p}' - \tfrac{63}{4} \mathit{p}'^2\right) \omega^2\bigr\} \Big]. \label{KillingCondition}
\end{align}
Contracting equation  \eqref{KillingCondition} with $\overset{*}{F}_{a}$ from \eqref{ForceVector012}, we arrive at the following scalar constraint equation,
\begin{equation}
H_{ab}\mathcal{F}^{ab} = 0.  \label{KillingCondition1}
\end{equation}
Where the tensor $\mathcal{F}^{ab}$ has been defined as follows,
\begin{align}
\mathcal{F}^{ab} &= \lambda^2 \left(\mathit{p}'^2 \dot{U}^{a} \dot{U}^{b} \Theta^2 -  Z^{a} Z^{b} \omega^2\right)\nonumber\\
&=\dot{u}^{a}\dot{u}^{a}  \Theta^{2}- \tilde{\nabla}^{a}\Theta\tilde{\nabla}^{b}\Theta.  \label{ShearForce}
\end{align} 
In equation \eqref{KillingCondition1}, it is clear that whenever  $\mathcal{F}^{ab}$ vanishes, then the equation \eqref{ShearForce} becomes $\tilde{\nabla}^{a}\Theta\tilde{\nabla}^{b}\Theta=(\Theta\dot{u}^{a})(\Theta\dot{u}^{b})$, which implies that $F^{a}$ or $\overset{*}{F}^{a}$ are zero, and this has been dealt with in the case where $\dot{U}^{a}\hat{\omega}_{a}\neq0$ above. It remains to show that one of the solutions of equation \eqref{KillingCondition1}  comes from the Killing equation, where the Killing vector is parallel to the vorticity. 
\subsection{The existence of the Killing vector along the vorticity}
Let a vector $K^{a}$ be a Killing vector along the vorticity with a generic scalar coefficient $\mathcal{K}$, be expressed in the following way, 
\begin{equation}
K^{a} = \hat{\omega}^{a} \mathcal{K}.  \label{KillingVector}
\end{equation}
This vector satisfies the Killing equation below,
\begin{equation}
\nabla^{(a}K^{b)}=0.  \label{KillingEquation1}
\end{equation}
Equation \eqref{KillingEquation1} can be written alternatively in terms of the magnetic part of the Weyl tensor $H^{ab}$ by substituting for $K^{a}$ as follows,
\begin{align}
0 &=\tilde{\nabla}^{(a}K^{b)} \\
0 &=\mathcal{K} \tilde{\nabla}^{(a}\hat{\omega}^{b)} + \hat{\omega}^{(a} \tilde{\nabla}^{b)}\mathcal{K}\\
0 &=\frac{\mathcal{K}}{\omega}\tilde{\nabla}^{(a}\omega^{b)} - \frac{1}{\omega}K^{(a}\tilde{\nabla}^{b)}\omega + \hat{\omega}^{(a} \tilde{\nabla}^{b)}\mathcal{K}\\
0 &=\frac{\mathcal{K}}{\omega}H^{ab} + \frac{1}{\omega}K^{(a}\tilde{\nabla}^{b)}\omega -  \hat{\omega}^{(a} \tilde{\nabla}^{b)}\mathcal{K} + 2\dot{u}^{(a} K^{b)}.
\end{align}
Hence the magnetic part of the Weyl tensor $H^{ab}$ is given by:
\begin{align}
H^{ab} &= \omega^{(a}\Bigl[\tfrac{1}{\mathcal{K}}\tilde{\nabla}^{b)}\mathcal{K} - 2\dot{u}^{b)}\Bigr] - \hat{\omega}^{(a} \tilde{\nabla}^{b)}\omega \label{KillingEquation2} \\
            & = \omega^{(a} \tilde{\nabla}^{b)}\zeta  \label{HabKilling}
\end{align}
Where in equation \eqref{HabKilling}, $\zeta$ is given by equation \eqref{ZetaS}. The scalar $\mathcal{K}$ and hence the Killing vector $K^{a}$ satisfies the following equations below,
\begin{align}
\dot{\mathcal{K}} &= \tfrac{1}{3} \mathcal{K} \Theta, \label{KillingScalarDot}\\
\dot{K}^{\langle a\rangle} &= \tfrac{1}{3}\Theta K^{a} \label{KillingVDot}
\end{align}
\begin{equation}
\tilde{\nabla}^{a}\mathcal{K} = \mathcal{K} \bigl(\tfrac{2}{3} \lambda \epsilon^{a}{}_{bc}\hat{\omega}^{b} Z^{c} -2 \dot{u}^{a}  -  \tfrac{1}{\omega}\tilde{\nabla}^{a}\omega\bigr). \label{KillingEquation3}
\end{equation}
Where equations \eqref{KillingScalarDot} and \eqref{KillingEquation3} result from contracting equation \eqref{KillingEquation1} with the velocity $u_{b}$ the vorticity $\omega_{b}$ respectively. Equations \eqref{KillingScalarDot} and \eqref{KillingEquation3} can be recast in the following way, where $\Sigma$ is given by equation \eqref{SigmaS} in the appendix,
\begin{align}
\dot{\Sigma} &= 3\Theta\bigl(\mathit{p}' - \tfrac{1}{9}\bigr), \label{SigmaDot}\\
\tilde{\nabla}^{a}\Sigma &= \tfrac{2}{3} \epsilon^{a}{}_{bc} \lambda\hat{\omega}^{b} Z^{c}. \label{DaSigma}
\end{align}
Considering equation \eqref{KillingEquation2}, it is clear that this equation is a solution to equation \eqref{KillingCondition1} since the tensor $\mathcal{F}^{ab}$ is orthogonal to the vorticity. Finally, to nail down the existence of the Killing vector given by equation \eqref{KillingVector} we need to show that the Killing equations are integrable. By taking the curl of equation \eqref{KillingEquation3} and applying the commutation relations for the expansion scalar $\Theta$ equation \eqref{SSCofS} together with equation \eqref{KillingEquation3} itself, we can show that,
\begin{align}
\epsilon^{a}{}_{bc} \tilde{\nabla}^{b}\tilde{\nabla}^{c}\mathcal{K} = \mathcal{K}\left[-2\left( \tfrac{1}{3} - 3 \mathit{p}'\right) \Theta \omega^{a} - 2\epsilon^{a}{}_{bc} \tilde{\nabla}^{b}\dot{u}^{c} - \tfrac{1}{\omega}\epsilon^{a}{}_{bc}  \tilde{\nabla}^{b}\tilde{\nabla}^{c}\omega\right],
\end{align}
 then applying the commutation relations for the energy density $\mu$ using equation \eqref{SSCofS} and the vorticity scalar $\omega$ also using equation \eqref{SSCofS} we obtain the following integrability condition,
\begin{equation}
\epsilon^{a}{}_{bc} \tilde{\nabla}^{b}\tilde{\nabla}^{c}\mathcal{K} = \tfrac{2}{3} \mathcal{K} \Theta \omega^{a} = 2\dot{\mathcal{K}}\omega^{a}.  \label{KillingIntegrability}
\end{equation}
Where equation \eqref{KillingIntegrability}, is the just the spatial commutation relation for any scalar field $
\mathcal{K}$ given by equation \eqref{SSCofS}. By time propagating \eqref{KillingEquation3} and taking spatial gradient of equation \eqref{KillingScalarDot}, one can show that the following equation is satisfied,
\begin{equation}
h^{a}{}_{b} \bigl[\tilde{\nabla}^{b}\mathcal{K}\bigr]^{\cdot} = \tilde{\nabla}^{a}\dot{\mathcal{K}} + \dot{u}^{a} \dot{\mathcal{K}} -  \tfrac{1}{3} \Theta \tilde{\nabla}^{a}\mathcal{K} + \epsilon^{a}{}_{cb} \omega^{c} \tilde{\nabla}^{b}\mathcal{K}.\label{TSDCofK}
\end{equation}
This is the evolution equation for the spatial gradient of any scalar field $\mathcal{K}$ given by equation \eqref{TSDCofS}. Therefore, there exist a Killing vector along the vorticity. The following theorem (Theorem 4) below, shows that for $\Theta\omega$ to be zero, whenever $\omega\neq0$, one requires that $\dot{K}^{\langle a\rangle}$ must vanish.\\
\newline
\textbf{Theorem 4:} \textit{If a shear-free perfect fluid obeying a \textbf{barotropic} equation of state is rotating and admits a Killing vector field $K^{a}$ along the vorticity, then $\Theta\omega=0$, if and only if $\dot{K}^{\langle a\rangle}=0$.}\\
\newline
\textit{Proof:} If we assume that $\Theta\omega=0$. Then it follows that $\omega\neq0=\Theta$ since the fluid is rotating, therefore equation \eqref{KillingVDot} implies that, 
\begin{align}
\dot{K}^{\langle a\rangle} &=0.
\end{align}
Since $K^{a}$ is a Killing vector  along the vorticity defined by equation \eqref{KillingVector}, we have shown the first part of the proof. Conversely, if the Killing vector field along the vorticity given by $K^{a}$ has the property  that $\dot{K}^{\langle a\rangle}=0$. Therefore, by equation \eqref{KillingVDot}, $\Theta$ must vanish. Hence $\Theta\omega=0$ and we have the proof. \newline

\textit{Remark:} The spatially projected Lie derivative of the vorticity vector along the velocity is given in the following way,
\begin{equation}
h^{a}{}_{b} \mathcal{L}_u \omega^{b} = \Theta \omega^{a}\left(\mathit{p}' - 1\right). \label{LieDOmega}
\end{equation}
In the spirit of causality, we know that $\mathit{p}' \leq1$, therefore we can rewrite equation \eqref{LieDOmega} as follows whenever $\mathit{p}'<1$,
\begin{equation}
h^{a}{}_{b}\mathcal{L}_u \omega^{b} = \dot{K}^{\langle a\rangle} \frac{3 \omega}{\mathcal{K}}\left(\mathit{p}' - 1\right).  \label{LieDOmega1}
\end{equation}
Hence, we have that whenever $\dot{K}^{\langle a\rangle} =0$ or $\mathit{p}'=1$, from the former it follows that $h^{a}{}_{b}\mathcal{L}_u \omega^{b}=0$, thus the vorticity vector must be a \textit{basic} vector field. In both cases we have that $\Theta\omega=0$, and we have the proof of the shear-free perfect fluid conjecture. Since the latter violates causality,  we must seek a causality obeying condition regrading the presence of the vorticity as in the following theorem.
\section{The existence of a \textit{basic} vector field along the direction of the vorticity}\label{Sec9}
In the presence of rotation, we have shown that if a Killing vector along the direction of the vorticity exists. Then theorem four above asserts that for the shear-free perfect fluid conjecture to be true, this Killing vector needs to have the property that $\dot{K}^{\langle a\rangle}=0$, which by theorem five below, implies the existence of a particular \textit{basic} vector field along the direction of the vorticity.\\\\
\textbf{Theorem 5:} \textit{If a shear-free perfect fluid obeying a \textbf{barotropic} equation of state is rotating, then $\Theta\omega=0$, if and only if there exist a \textbf{basic} vector field $\Omega^{a}=e^{\gamma}\omega^{a}$, with $\gamma=\int\tfrac{dp}{\mathcal{E}}$ along the direction of the vorticity.}\\
\newline
\textit{Proof:} A purely spatial vector field is said to be \textit{basic} if its spatially projected Lie derivative along $\bf u$ vanishes as defined in \cite{van2016shear}. \\
\newline
Therefore, the spatially projected Lie derivative along $\bf u$ of $\Omega^{a}$ is given by;
\begin{align}
h^{a}{}_{b} \mathcal{L}_u \Omega^{b} &= - e^{\gamma} \Theta \omega^{a}\label{BasicOmega} \\
                                     &= - \Theta\Omega^{a}. 
\end{align}
It is clear from equation \eqref{BasicOmega} that $\Omega^{a}$ is \textit{basic} if and only if $ \Theta \omega$ is zero. Alternatively, according to equation \eqref{KillingVector} and \eqref{KillingVDot}, equation \eqref{BasicOmega} can be written as   
\begin{equation}
h^{a}{}_{b} \mathcal{L}_u \Omega^{b} = - \frac{3 e^{\gamma}\omega}{\mathcal{K}}\dot{K}^{\langle a\rangle} ,\label{BasicOmega1}
\end{equation}
whenever a Killing vector exists along the direction of the vorticity. Therefore, $\Omega^{a}$ is \textit{basic} if  $\dot{K}^{\langle a\rangle}=0$, which by Theorem 4 above, this implies $\Theta\omega=0$, and we have the proof.\\
\newline
Conversely, assuming that $\Theta\omega=0$, thus $\Theta$ must vanish. Therefore, according to equation \eqref{BasicOmega} this implies that $h^{a}{}_{b} \mathcal{L}_u \Omega^{b} =0$. Hence $\Omega^{a}$ must be a \textit{basic} vector field, and we are done. \\

\textit{Remark:} Theorem five above implies that whenever the vorticity is non-zero, there exist at least a \textit{basic} vector along the direction of the vorticity, namely $\Omega^{a}$ apart from the Killing vector $K^{a}=\mathcal{K}\hat{\omega}^{a}$ and $\frac{\hat{\omega}^{a}}{\lambda}$ in order for the shear-free perfect fluid conjecture to be true. Taking the ratios of the scalar coefficients of these \textit{basic} vectors along the vorticity direction, we must obtain \textit{basic} functions as follows, 
\begin{equation}
\left[\frac{e^{\gamma} \omega}{\mathcal{K}}\right]^{.} = -\Theta \frac{e^{\gamma} \omega}{\mathcal{K}} = 0,
\end{equation}

\begin{equation}
\left[\lambda e^{\gamma} \omega\right]^{.} = \left[e^{\Xi} \omega\right]^{.} = -\Theta e^{\Xi} \omega = 0.
\end{equation}
Where $\Xi$ is given as follows,
\begin{equation}
\Xi=\int\tfrac{1}{\mathcal{E}}\left(p^{\prime}+\tfrac{1}{3}\right)d\mu. 
\end{equation}

\section{The existence of a \textit{basic} vector field along the direction of the acceleration}\label{Sec10}
It has been well established in section (\ref{Sec6}) by theorem one that whenever $\dot{u}^{a}=0$, the shear-free perfect fluid conjecture is true. The following theorem exhibits a condition by which the conjecture is satisfied in the presence of acceleration of a barotropic perfect fluid with $p(\mu)+\mu\neq0$. This theorem is similar in structure to theorem five above, and it is related to theorem one and two of reference  \cite{van2016shear}, but it is more general since it deals with a general covariant acceleration vector field instead of its individual components.\\\\
\textbf{Theorem 6:} {\it If the velocity vector field of a barotropic perfect fluid is shear-free, then either the expansion or the rotation vanishes if and only if there exist a \textbf{basic} vector field parallel to the acceleration vector field.}\\
\newline
\textit{Proof:} If the acceleration vector field is non-zero, we can define a vector field along it as follows,
\begin{equation}
\mathcal{V}^{a} = \frac{\dot{u}^{a}}{\lambda^2 \mathit{p}'}.\label{VaForUa}
\end{equation}
Where $\lambda=\exp\left[\int\frac{d\mu}{3\mathcal{E}}\right]$. The spatially projected time-propagation of $\mathcal{V}^{a}$ is given by,
\begin{equation}
\dot{\mathcal{V}}^{\langle a\rangle} = \frac{F^{a}}{\lambda^2} + \tfrac{1}{3} \mathcal{V}^{a} \Theta -  \epsilon^{a}{}_{cb} \mathcal{V}^{c} \omega^{b}. \label{VaForUaDot}
\end{equation}
For $\mathcal{V}^{a}$ to be a \textit{basic} vector field, it suffices to let its spatially projected Lie derivative along the velocity be zero. Hence, the spatially projected Lie derivative of $\mathcal{V}^{a}$ is given by,
\begin{align}
h^{a}{}_{b} \mathcal{L}_u \mathcal{V}^{b} &= \frac{F^{a}}{\lambda^2}\\
                                                                  &=\frac{\Theta}{\lambda^{2}}\tilde{\nabla}^{a}\xi. \label{LieDofVa}
\end{align}
Whenever $\mathcal{V}^{a}$ is \textit{basic}, then $F^{a}$ vanishes. We have shown in theorem three that whenever $F^{a}$ is zero, then $\Theta\omega=0$, and we have the proof.\\
\newline
Conversely, if $\Theta\omega=0$, we have two cases to consider. Case one where $\Theta=0$, and case two where $\omega=0$. We shall deal with these cases respectively in the following way, \\
\newline
\textbf{Case 1. } When $\Theta=0$, clearly by equation \eqref{LieDofVa} $h^{a}{}_{b} \mathcal{L}_u \mathcal{V}^{b}$ is zero. Therefore, $\mathcal{V}^{a}$ is a \textit{basic} vector field, and the proof is finished.\\
\newline
\textbf{Case 2. } When $\omega=0$, clearly by equation \eqref{eq:RICE:1} $\tilde{\nabla}^{a}\Theta$ is zero. Therefore, we have that $\nabla^{a}\Theta=-u^{a}\dot{\Theta}$, hence $u^{a}$ is hyper-surface orthogonal (obviously since $\omega=0$). This means we can write that $\nabla^{a}\xi=-u^{a}\dot{\xi}$, and therefore $\tilde{\nabla}^{a}\xi$ is zero. Which implies that $F^{a}$ must vanish, and by equation \eqref{LieDofVa} it follows that $h^{a}{}_{b} \mathcal{L}_u \mathcal{V}^{b}$ is zero. Hence $\mathcal{V}^{a}$ must be \textit{basic}, thus the proof is complete. 
\section{Discussion and Conclusions}\label{Sec11}
After setting up the formalism needed and displaying results that hold for any perfect fluid in GR, we specialize these results to the shear free case and then derive thirteen levels of constraint that must hold in this situation. We then prove some specific cases using a symbolic computer algebra package in Mathematica called xTensor~\cite{DropboxFolder}.  we have managed to reproduce two covariant proofs provided in reference \cite{senovilla1998theorems}, these are the constant pressure case and the case where the vorticity is parallel to the acceleration vector field, illuminating some of the steps involved in these proofs. We then provide covariant proofs of the cases where the vorticity is orthogonal to the acceleration and where there exist a \textit{basic} vector along the acceleration, which were first given in reference \cite{van2016shear} using a tetrad formalism.\\
\newline
In general $(\text{when}, \Theta\omega\neq0)$, if we contract equation \eqref{NewConstr13} with $ \overset{*}{F}^{a}$, we obtain the scalar constraint equation given in \eqref{NewC14}. Since $\omega\neq0$, there exist a Killing vector along the vorticity given by equation \eqref{KillingVector} . Then by applying the restrictions imposed by the existence of the Killing vector on the scalars \eqref{UScalar} to \eqref{HScalar}, equation \eqref{NewC14} is satisfied. This Killing vector changes in time by the amount $\tfrac{1}{3}\Theta$, therefore the shear-free perfect fluid conjecture is true if the Killing vector has the property that $\dot{K}^{\langle a\rangle}=0$, which essentially means that there exist a \textit{basic} vector field along the direction of the vorticity (Theorem 5). It seems that in the case of a generic barotropic equation of state, the shear-free conjecture is true only for special cases of the vorticity vector field (either the vorticity is zero or there is a \textit{basic} vector $\Omega^{a}$ along it).\\
\newline
Finally, we have shown that if the acceleration is not zero, the exist a \textit{basic} vector $\mathcal{V}^{a}$ along it for the conjecture to be true. We believe that Theorem 6 puts serious restrictions on shear-free perfect fluids in GR. The shear-free perfect fluid conjecture is true if and only if either the vorticity or the acceleration vectors are aligned with some hypersurface \textit{basic} vectors. We deduce that the general proof of the shear-free  perfect fluid conjecture will encompass the analysis based on \textit{basic} variables of the theory as can be seen from reference~\cite{van2016shear}. For the upcoming work,  we will use the techniques developed in this work to extend this analysis to the case of $f(R)$ gravity. In particular we will examine the stability of the GR version of this theorems by considering theories of the form $f(R)=R^{1+\delta}$ , where $\delta<<1$ as our starting point. 
\section*{Acknowledgments}
We would like to thank Prof. Norbert Van den Bergh for contributing the to the formal structure of the proofs in this work and for his critical comments, valuable discussions and for checking out our calculations.  And we are grateful to Prof. George Ellis for his introductory remarks on putting the work done here in perspective and being a source of inspiration for the search of the shear-free perfect fluid conjecture proof. I would also like to thank the support of the South African SKA bursary scheme for supporting this work through their support for my PhD. And also to Dr. Obinna Umeh for his help in introducing us to the Mathematica package xTensor which made most of this work possible.

\appendix
\section{}\label{Appendix}
 \subsection{Definations}
\begin{align}
\mathcal{U} &= \dot{U}^{a} \hat{\omega}_{a} \label{UScalar}\\
\mathcal{F} &= \frac{F^{a} \hat{\omega}_{a}}{\lambda}\label{FScalar}\\
\mathcal{G} &= -\mathcal{F} \lambda \left( \tfrac{1}{3} + \mathit{p}'\right) \omega + \mathcal{E} \lambda \mathit{p}^{\prime\prime} \mathcal{U} \Theta \omega \label{GScalar}\\
\mathcal{H} &= - \tfrac{1}{2} \mathcal{G} -  \tfrac{2}{9} \mathcal{F} \lambda \omega \label{HScalar}\\
\mathcal{I} &= - \mathcal{H} -  \tfrac{8}{9} \mathcal{F} \lambda \omega \label{IScalar}\\
\mathcal{X}& = - \tfrac{5}{3} + 2 \phi -  \frac{\mathcal{E}^2 \mathit{p}^{(3)}}{\mathit{p}'^2} + \frac{4}{9 \mathit{p}'} -  \frac{5 \phi}{3 \mathit{p}'} + \frac{\phi^2}{\mathit{p}'} + \mathit{p}' \label{XScalar}\\
\Sigma &= -2 \gamma + \log\left(\mathcal{K} \omega\right)  \label{SigmaS}\\
\zeta &= 2 \log(\mathcal{K}) -  \Sigma  \label{ZetaS}
\end{align}
\begin{equation} 
\omega^{a}{}_{c}\omega^{bc}=Q^{ab}\omega^2=\left(h^{ab}-\hat{\omega}^a\hat{\omega}^b\right)\omega^2 \label{ProjQab}
\end{equation} 
 \subsection{Constraint}
\begin{align}
0 &= \mathcal{F}^{ab} H_{ab} + \overset{*}{F}^{a} \dot{u}_{a} \Bigl[9 \mathcal{G} + \tfrac{9}{2} \mathcal{H} - \tfrac{1}{8} \mathcal{F} \lambda \bigl(1 - 14 \phi - 45 \mathit{p}'\bigr) \omega +  \tfrac{1}{2} \lambda \mathit{p}' \bigl(6 \nonumber \\ 
&\qquad - 7 \phi - 18 \mathit{p}'\bigr) \mathcal{U} \Theta \omega\Bigr] + \mathcal{F} \Bigl[\tfrac{1}{24} \mathcal{J} \lambda (10 + 21 \phi) \Theta \omega - \tfrac{1}{4} \mathcal{E} \lambda \bigl(1 + 3 \mathit{p}'\bigr) \Theta \omega\nonumber \\ 
&\qquad - \tfrac{1}{8} \lambda \dot{u}^2 (12 + 8 \phi - 36 \mathit{p}' - 7 \mathcal{X}) \Theta \omega + \tfrac{1}{24} \lambda \bigl(71 - 90 \mathit{p}' - 189 \mathit{p}'^2\bigr) \Theta \omega^3\Bigr] \nonumber \\ 
&\qquad + \mathcal{U} \Bigl[\tfrac{1}{8} \mathcal{F}^{a}{}_{a} \lambda \mathit{p}' \bigl(-13 + 9 \mathit{p}'\bigr) \omega - \tfrac{1}{12} \mathcal{J} \lambda (10 + 21 \phi) \mathit{p}' \Theta^2 \omega + \tfrac{1}{2} \mathcal{E} \lambda \mathit{p}' \bigl(1 \nonumber \\ 
&\qquad + 3 \mathit{p}'\bigr) \Theta^2 \omega +  \tfrac{1}{4} \lambda \mathit{p}' \dot{u}^2 (12 + 8 \phi - 36 \mathit{p}' - 7 \mathcal{X}) \Theta^2 \omega - \tfrac{1}{12} \lambda \mathit{p}' \bigl(71 - 90 \mathit{p}'\nonumber \\ 
&\qquad - 189 \mathit{p}'^2\bigr) \Theta^2 \omega^3\Bigr] + \tfrac{9}{4} \overset{*}{F}^{a} \left(\tilde{\nabla}_{a}\mathcal{G} + \tilde{\nabla}_{a}\mathcal{H}\right). \label{NewC14}
\end{align}

 \subsection{Propagation Equations}
 \begin{equation}
u^{c} \nabla_{c}\dot{U}^{a} = \lambda \mathit{p}' u^{a} \dot{U}_{c} \dot{U}^{c} + \mathit{p}' \dot{U}^{a} \Theta -  \epsilon^{a}{}_{cb} \dot{U}^{c} \hat{\omega}^{b} \omega + Z^{a} \omega  \label{UaDot}
\end{equation}
\begin{equation}
h^{a}{}_{b} \mathcal{L}_u V^{b} = \dot{V}^{\langle a\rangle} - \tfrac{1}{3} V^{a} \Theta + \epsilon^{a}{}_{bc} V^{b} \omega^{c} \label{ProjLieD}
\end{equation}
\subsection{Commutation relations}
\subsubsection{Time-space and space-space derivative commutator of scalars}
\begin{equation}
h^{a}{}_{b} \bigl[\tilde{\nabla}^{b}S\bigr]^{\cdot} = \tilde{\nabla}^{a}\dot{S} + \dot{u}^{a} \dot{S} -  \tfrac{1}{3} \Theta \tilde{\nabla}^{a}S + \epsilon^{a}{}_{cb} \omega^{c} \tilde{\nabla}^{b}S \label{TSDCofS}
\end{equation}
\begin{equation}
\tilde{\nabla}^{a}\tilde{\nabla}^{b}S  = \tilde{\nabla}^{b}\tilde{\nabla}^{a}S + 2 \epsilon^{ab}{}_{c} \omega^{c}\dot{S} \label{SSCofS}
\end{equation}

\subsubsection{Spatial gradients of 3-vectors: Time-space derivative commutators.}
\begin{align}
h^{a}{}_{e} h^{b}{}_{d}\bigl[\tilde{\nabla}^{e}\omega^{d}\bigr]^{\cdot} &= \tilde{\nabla}^{a}\dot{\omega}^{\langle b\rangle} + \dot{u}^{a} \dot{\omega}^{\langle b\rangle} -  \tfrac{1}{3} \Theta \tilde{\nabla}^{a}\omega^{b} + \epsilon^{a}{}_{cd} \omega^{c} \tilde{\nabla}^{d}\omega^{b} + \epsilon^{ab}{}_{d}\omega^{d}\dot{u}^{c} \omega_{c} \nonumber \\ 
&\qquad  + \bigl[(C_3)^{ad} -  H^{ad}\bigl] \epsilon^{b}{}_{cd}\omega^{c}  + \omega^{a}\bigl[\tfrac{1}{2} (C_1)^{b} -  \tfrac{1}{3} \Theta\dot{u}^{b}\bigr] \nonumber \\ 
&\qquad - h^{ab} \bigl[ \tfrac{1}{2} (C_1)^{c} - \tfrac{1}{3} \Theta\dot{u}^{c}\bigr] \omega_{c} \label{TSDCofOmega}
\end{align}
\begin{align}
h^{a}{}_{e} h^{b}{}_{d} \bigl[\tilde{\nabla}^{e}\dot{u}^{d}\bigr]^{\cdot} &= \tilde{\nabla}^{a}\ddot{u}^{\langle b\rangle} -  \tfrac{1}{3} \Theta \tilde{\nabla}^{a}\dot{u}^{b} + \epsilon^{a}{}_{cd} \bigl(\omega^{c} \tilde{\nabla}^{d}\dot{u}^{b} - \dot{u}^{b} \dot{u}^{c} \omega^{d}\bigr) + \epsilon^{ab}{}_{d} \omega^{d}\dot{u}_{c} \dot{u}^{c} \nonumber \\ 
&\qquad + \bigl[(C_3)^{ad} -  H^{ad}\bigr] \epsilon^{b}{}_{cd}\dot{u}^{c} + \dot{u}^{a} \bigl[\tfrac{1}{2} (C_1)^{b} + \ddot{u}^{\langle b\rangle} -  \tfrac{1}{3} \Theta\dot{u}^{b}\bigr] \nonumber\\
&\qquad - h^{ab}  \bigl[ \tfrac{1}{2} (C_1)^{c} - \tfrac{1}{3} \Theta\dot{u}^{c} \bigr]\dot{u}_{c} \label{TSDCofAcc}
\end{align}
{\it Space-space derivative commutators:}
\begin{align}
\tilde{\nabla}^{[a}\tilde{\nabla}^{b]}\omega_{c} &= E^{[a}{}_{c} \omega^{b]} + h^{[a}{}_{c} E^{b]}{}_{d}\omega^{d} + \tfrac{1}{3}( \Lambda -  \tfrac{1}{3} \Theta^2 +  \mu) h^{[a}{}_{c}\omega^{b]} \nonumber \\ 
&\qquad  - \tfrac{1}{3} \Theta\epsilon^{[a}{}_{cd} \omega^{b]} \omega^{d} + \epsilon^{ab}{}_{d} \omega^{d} \dot{\omega}_{\langle c\rangle} \label{SSDCofOmega}
\end{align}
\begin{equation}
\tilde{\nabla}_{[a}\omega_{b]} = \tfrac{1}{2} \epsilon_{abc}(C_1)^{c} -  2\dot{u}_{[a} \omega_{b]} + \tfrac{1}{3} \epsilon_{abc} \tilde{\nabla}^{c}\Theta \label{GradComOmega}
\end{equation}
\begin{align}
\tilde{\nabla}^{[a}\tilde{\nabla}^{b]}\dot{u}_{c} &= E^{[a}{}_{c} \dot{u}^{b]} + h^{[a}{}_{c} E^{b]}{}_{d} \dot{u}^{d} + \tfrac{1}{3}\bigl( \Lambda -  \tfrac{1}{3} \Theta^2 + \mu -  3\omega^2\bigr)h^{[a}{}_{c} \dot{u}^{b]} \nonumber\\
&\qquad  -  \tfrac{1}{3} \Theta\epsilon^{[a}{}_{cd} \dot{u}^{b]} \omega^{d} -  \tfrac{1}{3}\Theta h^{[a}{}_{c}\epsilon^{b]}{}_{de} \dot{u}^{d} \omega^{e} - \dot{u}^{[a} \omega^{b]} \omega_{c} \nonumber \\ 
&\qquad  + \dot{u}^{d}\omega_{d} h^{[a}{}_{c}\omega^{b]} + \epsilon^{ab}{}_{d}\omega^{d} \ddot{u}_{\langle c\rangle} \label{SSDCofAcc}
\end{align}
\begin{equation}
\tilde{\nabla}_{[a}\dot{u}_{b]} = \epsilon_{abc}\omega^{c}\mathit{p}' \Theta \label{GradComAcc}
\end{equation}

\subsubsection{Spatial gradients of 3-tensors: Evolution of spatial divergence terms}
\begin{align}
h^{a}{}_{b} \bigl[\tilde{\nabla}_{c}H^{bc}\bigr]^{\cdot} &= \tilde{\nabla}_{b}\dot{H}^{\langle ab\rangle} + \dot{u}_{b}\dot{H}^{\langle ab\rangle} - \tfrac{1}{3} \Theta \tilde{\nabla}_{b}H^{ab} - \epsilon_{bcd} \omega^{b} \tilde{\nabla}^{d}H^{ac} \nonumber \\ 
&\qquad + \Theta \dot{u}_{b}H^{ab} + \epsilon_{bcd} H^{ad} \dot{u}^{b} \omega^{c} + \epsilon^{a}{}_{cd} H_{b}{}^{d} \dot{u}^{b} \omega^{c} \nonumber \\ 
&\qquad - \tfrac{3}{2} H^{a}{}_{b}(C_1)^{b}  - \epsilon^{a}{}_{cd} (C_3)^{bc}H_{b}{}^{d}  \label{TSDCofDivH}
\end{align}
\begin{align}
h^{a}{}_{b}\bigl[\tilde{\nabla}_{c}E^{bc}\bigr]^{\cdot} &= \tilde{\nabla}_{b}\dot{E}^{\langle ab\rangle} + \dot{u}_{b}\dot{E}^{\langle ab\rangle} -  \tfrac{1}{3} \Theta \tilde{\nabla}_{b}E^{ab} -  \epsilon_{bcd} \omega^{b} \tilde{\nabla}^{d}E^{ac} - \epsilon^{a}{}_{cd} H_{b}{}^{d}E^{bc}  \nonumber \\ 
&\qquad  + \Theta\dot{u}_{b} E^{ab} + \epsilon_{bcd}E^{ad}\dot{u}^{b} \omega^{c} + \epsilon^{a}{}_{cd}E_{b}{}^{d} \dot{u}^{b} \omega^{c} - \tfrac{3}{2} E^{a}{}_{b}(C_1)^{b} \nonumber \\ 
&\qquad  -  \epsilon^{a}{}_{cd}(C_3)^{bc} E_{b}{}^{d} \label{TSDCofDivE}
\end{align}
{\it Space-space derivative commutators:}
\begin{align}
\tilde{\nabla}^{[a}\tilde{\nabla}^{b]}H_{cd} &= 2E^{[a}{}_{(c} H^{b]}{}_{d)} + 2h^{[a}{}_{(c}E^{b]e} H_{d)e} + \tfrac{2}{3}h^{[a}{}_{(c} H^{b]}{}_{d)} \bigl( \Lambda -  \tfrac{1}{3} \Theta^2 + \mu - 3\omega^2\bigr) \nonumber \\ 
&\qquad -  2 H^{[a}{}_{(c} \omega^{b]} \omega_{d)} -  \tfrac{2}{3} \Theta \epsilon^{[a}{}_{(ce}\omega^{e} H^{b]}{}_{d)} + \tfrac{2}{3}\Theta h^{[a}{}_{(c}\epsilon^{b]}{}_{ej} \omega^{e}H_{d)}{}^{j} \nonumber \\ 
&\qquad + 2 h^{[a}{}_{(c}\omega^{b]} H_{d)e} \omega^{e} + \epsilon^{ab}{}_{e}\omega^{e} \dot{H}_{\langle cd\rangle} \label{SSDCofHab}
\end{align}
\begin{align}
\tilde{\nabla}^{[a}\tilde{\nabla}^{b]}E_{cd} &= -  2 E^{[a}{}_{(c} \omega^{b]} \omega_{d)} + 2h^{[a}{}_{(c}E^{b]e} E_{d)e} + \tfrac{2}{3}h^{[a}{}_{(c}E^{b]}{}_{d)} \bigl( \Lambda -  \tfrac{1}{3} \Theta^2 + \mu - 3 \omega^2\bigr) \nonumber \\ 
&\qquad + 2 h^{[a}{}_{(c} \omega^{b]}E_{d)e}\omega^{e} -  \tfrac{2}{3} \Theta\epsilon^{[a}{}_{(ce} \omega^{e}E^{b]}{}_{d)} + \tfrac{2}{3} \Theta h^{[a}{}_{(c}\epsilon^{b]}{}_{ej}\omega^{e}E_{d)}{}^{j}\nonumber\\
&\qquad + \epsilon^{ab}{}_{e} \omega^{e}\dot{E}_{\langle cd\rangle} \label{SSDCofEab}
\end{align}

\section{}
The following algebraic and differential identities have been applied in the time propagation of the constraint equations in this work, see reference \cite{maartens1997linearization,van1996extensions}. 
\subsection{Identities}

\begin{equation}
0 =  \epsilon^{abc}\tilde{\nabla}_{\langle b}\omega_{d\rangle} \dot{u}_{c} \omega^{d} - \epsilon^{abc}\tilde{\nabla}_{\langle b}\omega_{d\rangle} \dot{u}^{d} \omega_{c} - \tilde{\nabla}^{\langle a}\omega^{b\rangle} \epsilon_{bcd} \dot{u}^{c} \omega^{d}\label{IdentDivC10}
\end{equation}
\begin{equation}
0 = \epsilon^{a}{}_{bd} \dot{u}^{b} \tilde{\nabla}_{c}H^{cd} -  \epsilon_{bcd} \dot{u}^{b} \tilde{\nabla}^{d}H^{ac} + \epsilon^{a}{}_{cd} \dot{u}^{b} \tilde{\nabla}^{d}H_{b}{}^{c} \label{Ident2C4}
\end{equation}
\begin{equation}
0 = \epsilon^{a}{}_{bd} \omega^{b} \tilde{\nabla}_{c}E^{cd} -  \epsilon_{bcd} \omega^{b} \tilde{\nabla}^{d}E^{ac} + \epsilon^{a}{}_{cd} \omega^{b} \tilde{\nabla}^{d}E_{b}{}^{c} \label{Ident3C4}
\end{equation}
\begin{equation}
0 = - \epsilon_{bcd} H^{ad} \tilde{\nabla}^{c}\dot{u}^{b} + \epsilon^{a}{}_{cd} H_{b}{}^{d} \tilde{\nabla}^{c}\dot{u}^{b} -  \epsilon^{a}{}_{bd} H_{c}{}^{d} \tilde{\nabla}^{c}\dot{u}^{b} \label{Ident5bC4}
\end{equation}
\begin{equation}
0 = E_{c}{}^{d} \epsilon^{a}{}_{bd} \tilde{\nabla}^{c}\omega^{b} -  E_{b}{}^{d} \epsilon^{a}{}_{cd} \tilde{\nabla}^{c}\omega^{b} + E^{ad} \epsilon_{bcd} \tilde{\nabla}^{c}\omega^{b} \label{Ident6C4}
\end{equation}
\begin{equation}
0 = - E_{c}{}^{d} \epsilon^{a}{}_{bd} \dot{u}^{b} \omega^{c} + E_{b}{}^{d} \epsilon^{a}{}_{cd} \dot{u}^{b} \omega^{c} -  E^{ad} \epsilon_{bcd} \dot{u}^{b} \omega^{c} \label{Ident8C4}
\end{equation}
\begin{equation}
0 = \epsilon^{a}{}_{bd} \dot{u}^{b} \tilde{\nabla}_{c}E^{cd} -  \epsilon_{bcd} \dot{u}^{b} \tilde{\nabla}^{d}E^{ac} + \epsilon^{a}{}_{cd} \dot{u}^{b} \tilde{\nabla}^{d}E_{b}{}^{c} \label{ident1C5}
\end{equation}
\begin{equation}
0 = \epsilon^{a}{}_{bd} \omega^{b} \tilde{\nabla}_{c}H^{cd} -  \epsilon_{bcd} \omega^{b} \tilde{\nabla}^{d}H^{ac} + \epsilon^{a}{}_{cd} \omega^{b} \tilde{\nabla}^{d}H_{b}{}^{c} \label{ident2C5}
\end{equation}
\begin{equation}
0 = \epsilon_{bcd} H^{ad} \tilde{\nabla}^{c}\omega^{b} -  \epsilon^{a}{}_{cd} H_{b}{}^{d} \tilde{\nabla}^{c}\omega^{b} + \epsilon^{a}{}_{bd} H_{c}{}^{d} \tilde{\nabla}^{c}\omega^{b} \label{ident8C5}
\end{equation}
\begin{equation}
0 = - \epsilon_{bcd} H^{ad} \dot{u}^{b} \omega^{c} + \epsilon^{a}{}_{cd} H_{b}{}^{d} \dot{u}^{b} \omega^{c} -  \epsilon^{a}{}_{bd} H_{c}{}^{d} \dot{u}^{b} \omega^{c} \label{Ident2}
\end{equation}
\begin{equation}
0 = - E_{c}{}^{d} \epsilon^{a}{}_{bd} \tilde{\nabla}^{c}\dot{u}^{b} + E_{b}{}^{d} \epsilon^{a}{}_{cd} \tilde{\nabla}^{c}\dot{u}^{b} -  E^{ad} \epsilon_{bcd} \tilde{\nabla}^{c}\dot{u}^{b} \label{ident11C5}
\end{equation}

\section*{References}
\bibliography{iopart-num}

\end{document}